\documentclass[aps,prl,twocolumn,footinbib,amsmath,amssymb,amsfonts]{revtex4-2}

\usepackage{color,framed,graphicx}
\usepackage{tikz}

\newcommand{\HS}[1]{\textcolor{black}{{#1}}}
\newcommand{\JS}[1]{\textcolor{black}{{#1}}}
\newcommand{\DHK}[1]{\textcolor{black}{{#1}}}

\begin{document}
 
\title{Universal Thermal Entanglement of Multichannel Kondo Effects}

\author{Donghoon Kim} \email[]{These authors contributed equally to this work.}
\affiliation{Department of Physics, Korea Advanced Institute of Science and Technology, Daejeon 34141, Korea}

\author{Jeongmin Shim} \email[]{These authors contributed equally to this work.}
\affiliation{Department of Physics, Korea Advanced Institute of Science and Technology, Daejeon 34141, Korea}

\author{H.-S. Sim}\email[]{hssim@kaist.ac.kr}
\affiliation{Department of Physics, Korea Advanced Institute of Science and Technology, Daejeon 34141, Korea}

\date{\today}

\begin{abstract}
Quantum entanglement between an impurity and its environment is expected to be central in quantum impurity problems. We develop a method to compute the entanglement in spin-$1/2$ impurity problems, based on the entanglement negativity and the boundary conformal field theory (BCFT).  Using the method, we study the thermal decay of the entanglement in the multichannel Kondo effects. At zero temperature, the entanglement 
has the maximal value independent of the number of the screening channels. At low temperature, the entanglement exhibits a power-law thermal decay.  The power-law exponent equals two times of the scaling dimension of the BCFT boundary operator describing the impurity spin, and it is attributed to the energy-dependent scaling behavior of the entanglement in energy eigenstates.
 These agree with numerical renormalization group results, unveiling quantum coherence inside the Kondo screening length.
\end{abstract}
\maketitle


{\it Introduction.---}
Quantum entanglement has been used to identify and characterize many-body states~\cite{Amico08,Eisert10,Laflorencie16,Wen17}.
It provides fundamental understanding especially when states possess maximal entanglement such as Bell entanglement~\cite{Affleck87,Kitaev01,Kitaev03}.
An interesting direction is to see how entanglement changes as a state deviates from a fixed point, e.g., thermally. Entanglement in pure excited states or thermal states has not been much studied~\cite{Alba09,Calabrese15,Park17}.

\begin{figure}[b]
\centerline{\includegraphics[width=.5\textwidth]{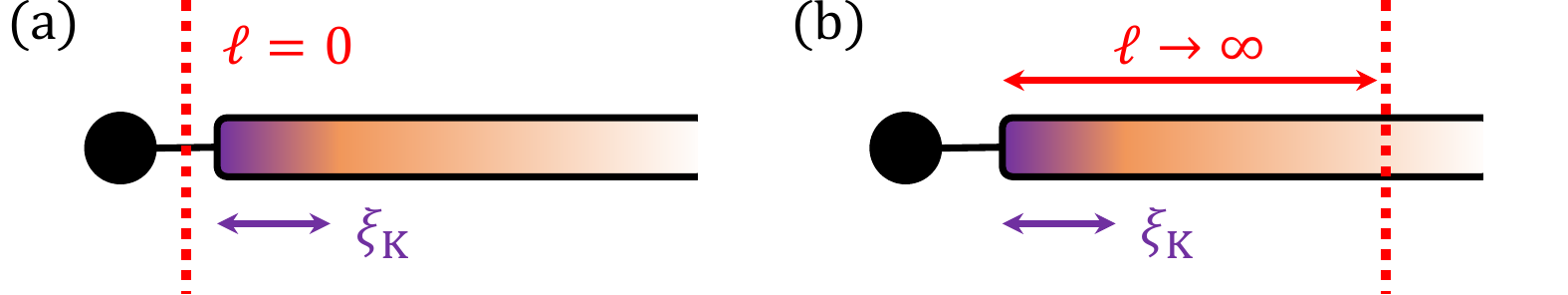}}
\caption{Bipartition (dotted lines) for entanglement in a Kondo system.
(a) It separates the impurity (circle) and the screening channels (rod).
(b) It is located at distance $l$ from the impurity far outside the Kondo cloud length $\xi_K$.
}
\label{MCKE_fig1}
\end{figure}

For this direction, there are a wide class of states in quantum impurity problems~\cite{Affleck08}, 
including impurities in metals~\cite{Kondo64,Hewson97}, spin chains~\cite{Bayat10a,Bayat12},  Luttinger liquids~\cite{Oshikawa06}, and quantum Hall effects~\cite{Fendley06,Fiete08}.
Here bipartite entanglement [see Fig.~\ref{MCKE_fig1}(a)] between an impurity and its environment will be central.
This entanglement was studied in the single-channel Kondo effect (1CK). In the 1CK ground state~\cite{Yoshida66}, it is Bell entanglement and induces impurity-spin screening. Its thermal suppression,  computed with numerical renormalization group methods (NRG)~\cite{Lee15,Shim18}, shows Fermi liquid behavior~\cite{Nozieres78,Mora09,Karki18}. 
The entanglement is used~\cite{Yoo18} for quantifying spatial distribution of Kondo clouds~\cite{ Affleck09b,Affleck01,Mitchell11,Park13,Borzenets20}.
It is valuable to study the entanglement in other impurities, including multichannel Kondo effects that show non-Fermi liquids~\cite{Nozieres80}, 
boundary phase transitions~\cite{Vojta06}, and fractionalization.
The entanglement will have essential information about how boundary degrees of freedom 
quantum-coherently couple with the bulk 
in boundary critical phenomena~\cite{Cardy89,Cardy91}.
 
However no approach for analytically computing the entanglement 
has been developed.  In a boundary conformal field theory (BCFT)~\cite{Affleck91b,Affleck93,Ludwig94}, a standard theory for quantum impurities and multichannel Kondo effects, 
the entanglement was not considered, 
since the impurity degrees of freedom are replaced by a boundary condition of the environment and
the entanglement is for the partition inside the Kondo cloud length.
By contrast,  
entanglement for another partition [Fig.~\ref{MCKE_fig1}(b)] far outside the Kondo length has been extensively computed~\cite{Sorensen07,Affleck09a,Eriksson11,Alkurtass16,Cornfeld17}, revealing the ``fractional ground-state degeneracy''~\cite{Affleck91a}.



Another difficulty arises 
in studying the entanglement in thermal states.
Entanglement entropy, a widely-used entanglement measure, cannot distinguish the entanglement from classical correlations~\cite{Plenio07,Guhne09,Horodecki09} in the mixed states, overestimating the entanglement.
Entanglement negativity~\cite{Lee00,Vidal02,Plenio05} is then a good choice, as it is applicable to mixed states.
This entanglement measure has been numerically computed for Kondo systems~\cite{Bayat10a,Alkurtass16,Shim18}.
 
In this work we develop an approach for analytically computing entanglement negativity $\mathcal{N}_\textrm{I$|$E}$ for the partition [Fig.~\ref{MCKE_fig1}(a)] between the impurity and its environment in a one-dimensional spin-1/2 impurity problem described by a BCFT.
The impurity part, replaced by a boundary condition in the BCFT, is restored in low-energy eigenstates, 
by identifying the impurity spin with BCFT boundary operators and computing its matrix elements with respect to the eigenstates.
Then the thermal density matrix is constructed, to obtain $\mathcal{N}_\textrm{I$|$E}$.
 
We then analyze thermal decay of the negativity in the $k$-channel Kondo effects ($k$CK).
At zero temperature, the entanglement between the impurity and screening channels
is maximal, $\mathcal{N}_\textrm{I$|$E}=1$, regardless of the number $k$ of the channels, although frustration in the impurity screening occurs $k$-dependently.
This is in stark contrast with the $k$-dependent 
impurity entropy for the partition in Fig.~\ref{MCKE_fig1}(b). At low temperature $T  \ll T_{\text{K}}$, the entanglement has algebraic thermal decay 
\begin{align}\label{scalingT}
\mathcal{N}_\textrm{I$|$E}(T) = 1 - a_{k} (\frac{T}{T_\text{K}})^{2\Delta}
\end{align}
with the exponent  identical to two times of the scaling dimension $\Delta$ of the BCFT operator describing the impurity spin. 
$\Delta = 1$ for $k=1$, $\Delta = 2/(2+k)$ for $k \ge 2$, $T_\text{K}$ is the Kondo temperature, and $a_{k}$ is a $k$-dependent positive constant. 
We confirm Eq.~\eqref{scalingT}, using the NRG method developed in Ref.~\cite{Shim18}.
We also \HS{find the scaling behavior of $\mathcal{N}_\textrm{I$|$E}$ in thermal crossover between different fixed points in channel anisotropic Kondo effects.}

Interestingly, the power-law exponent of the thermal decay \HS{is determined by that of the energy dependence of the entanglement} in energy eigenstates.
\HS{This is valid for general spin-1/2 impurities described by BCFTs.}

{\it \HS{Computation of negativity}.---}
We develop a method of computing the negativity \HS{ $\mathcal{N}_\textrm{I$|$E} = \Vert \rho^{\text{T}_\text{I}} \Vert - 1$} between the impurity and the other part in 
a one-dimensional critical system having a spin-1/2 impurity described by a BCFT at temperature $T$.
$\rho$ is the density matrix of the whole system, 
$\Vert \cdot \Vert$ is the trace norm, and \HS{$\text{T}_\text{I}$ is the partial transpose on the impurity}.
\DHK{The maximum possible value of $\mathcal{N}_\textrm{I$|$E}$ is 1 in the system~\cite{supp}.}

We organize steps for computing $\mathcal{N}_\textrm{I$|$E}$.
(i) At a fixed point, the impurity spin $\vec{S}_\text{imp}$ is identified with BCFT boundary operators $\psi_\alpha$ with scaling dimension $\Delta_\alpha$, 
\begin{equation} \label{general_scaling1}
S_\text{imp}^{\alpha = x,y,z} = c_\alpha + d_\alpha \psi_\alpha + \cdots
\end{equation} 
with constants $c_\alpha$, $d_\alpha$ and operators $\cdots$ of dimension $> \Delta_\alpha$. 
(ii) The energy $E$ ($\sim E_i, E_j$) dependence of matrix elements  $\langle E_{i} | \vec{S}_{\text{imp}} |E_{j} \rangle$ of the identified operator is studied \HS{for} low-energy eigenstates $ |E_{i} \rangle$ of the BCFT Hamiltonian, which includes the irrelevant terms describing thermal deviation from the fixed point. 
For the purpose, we consider a finite system size $L$ and choose $L \sim v / E$, where $v$ is a relevant velocity (we set the Planck constant $\hbar \equiv 1$ and Boltzmann constant $k_\textrm{B} \equiv 1$).
\HS{Using conformal transformation~\cite{supp}, we find}
\begin{equation} 
\langle E_{i} | S_\text{imp}^\alpha |E_{j} \rangle = c_\alpha \delta_{ij} + d_\alpha O(L^{-\Delta_\alpha}),  \label{general_scaling2}
\end{equation}
and their energy dependence is obtained by replacing $v/L$ by $E$.
The replacement has been justified~\cite{Wilson75,Anders05,Weichselbaum07,Bulla08,Zarand00,Affleck91b,Affleck92} such that the states of energy $E$ in an infinite-size system, e.g., obtained with the NRG, are well described by the corresponding BCFT or bosonization of finite size $L \sim v/E$; the region outside $L$ negligibly affects the states of energy $E$ at positions near the impurity.
\HS{It is because correlations exponentially decay with distance $x \gtrsim v/T$ at temperature $T$.} The replacement allows us to avoid numerical calculations~\cite{Besken20} of the matrix elements.



  

(iii) The energy eigenstates $|E_{i} \rangle$'s are \HS{represented~\cite{supp} in the bipartite basis states of the impurity and the environment, utilizing Eq.~\eqref{general_scaling2}. The impurity degrees of freedom are restored in the representation. 
Energy dependence in the representation of $|E_{i} \rangle$ is found as in Eq.~\eqref{stateT} for the $k$CK.} 

\HS{Utilizing Schmidt decomposition~\cite{supp}, 
we find that each energy eigenstate $|E_{i} \rangle$ has the entanglement of
\begin{align}\label{purestate_negativity1}
\mathcal{N}_\textrm{I$|$E}(|E_i \rangle) = \sqrt{1 - 4 \langle E_i | \vec{S}_\textrm{imp} | E_i \rangle^2}
\end{align}
between the impurity and the environment, showing that the entanglement directly relates to the expectation value of the impurity spin for each low-energy engenstate. \DHK{This is a general relation applicable to any spin-1/2 impurities.}
Using Eq.~\eqref{general_scaling2} and expanding Eq.~\eqref{purestate_negativity1} upto possible leading contributions in the low-energy regime, we obtain 
\begin{align}
\mathcal{N}_\textrm{I$|$E} (| E_i \rangle) & = \sqrt{1 - 4 \sum_{\alpha=x,y,z} c^2_\alpha} + \sum_{\alpha} c_\alpha d_\alpha O (E_i^{\Delta_\alpha})  \nonumber \\
 & +  \sum_{\alpha} d_\alpha^2 O (E_i^{2 \Delta_\alpha}) + \cdots.
 \label{purestate_negativity3} \end{align}
The energy dependence of $\mathcal{N}_\textrm{I$|$E} (| E_i \rangle)$ follows $E_i^{\Delta_\alpha}$ or  $E_i^{2 \Delta_\alpha}$, depending on $c_\alpha$'s.
}
  
\HS{(iv)  The} thermal density matrix \HS{is constructed},  $\rho(T) = \sum_i w_{i} |E_{i} \rangle \langle E_{i}|$, \HS{with the eigenstates $|E_i \rangle$ of energy $E_i \sim T$, the Boltzmann weight $w_{i} = e^{-E_{i} / T}/Z$, and the partition function $Z = \sum_i e^{-E_{i}/T}$.}
This approximate density matrix was proved~\cite{Weichselbaum07,Bulla08} to well describe thermodynamic properties associated with the impurity; the exact density matrix is governed mainly by the eigenstates $|E_i \sim T \rangle$, \HS{because}
$w_i$ decays exponentially with $E_i \gtrsim T$ while the density of states increases with $E_i$. 
Then the negativity $\mathcal{N}_\textrm{I$|$E} (\rho) = \Vert \rho^{\text{T}_\text{I}} \Vert - 1$ is computed.

\HS{
Combining these steps, 
we find~\cite{supp} that the thermal behavior of the entanglement satisfies
\begin{align}
\label{purestate_negativity2}
\mathcal{N}_\textrm{I$|$E} (\rho(T)) = \sum_i w_i (E_i) \mathcal{N}_\textrm{I$|$E} (| E_i \rangle)|_{E_i \sim T} + f(T).
\end{align}
The first term of $\mathcal{N}_\textrm{I$|$E} (| E_i \sim T \rangle)$ is the leading contribution from the diagonal elements of $\rho$, while $f(T)$ is the other contribution from the diagonal and off-diagonal elements.
The first term is dominant at low temperature,
since $w_i(E_i \sim T) \sim O(1)$, $f(T) \sim T^\kappa$, and $\kappa$ \DHK{is larger than or equal to the minimum among $2 \Delta_\alpha$'s}.
Therefore the power-law exponent of the thermal behavior of $\mathcal{N}_\textrm{I$|$E} (\rho)$ equals that of the energy dependence of $\mathcal{N}_\textrm{I$|$E}(|E_i \rangle)$. 
Namely, the temperature dependence of $\mathcal{N}_\textrm{I$|$E}(\rho)$ stems from the universal behavior of the pure energy eigenstates.
This is a fixed-point property of general spin-$1/2$ impurities described by BCFTs. We below apply the findings to the $k$CK model.
}

{\it Restoring the impurity state in BCFT.---}
In the $k$CK model,  a spin-1/2 impurity $\vec{S}_\text{imp}$ interacts with $k$ channels of noninteracting electrons. Its Hamiltonian is
\begin{align}
H_{k\text{CK}} = \sum_{i=1}^k \Big[H_i + \lambda_i \vec{S}_\text{imp} \cdot \vec{S}_i \Big]
\end{align}
with the $i$th-channel Hamiltonian $H_i$, the interaction strength $\lambda_i>0$, and the $i$th-channel electron spin $\vec{S}_i$ at the impurity position.
We first consider isotropic couplings $\lambda_1=\cdots=\lambda_k = \lambda$ at $T \ll T_\textrm{K}$. This regime is described by the BCFT Hamiltonian~\cite{Affleck91b,Affleck93,Ludwig94,Affleck90} 
\begin{equation}
H_\textrm{BCFT} = H_\text{FP} + \bar{\lambda} H_\text{LI}.
\label{BCFTH}
\end{equation}
$H_\text{FP}$ is the fixed-point Hamiltonian invariant under $\mathrm{U}(1) \times \mathrm{SU}(2)_{k} \times \mathrm{SU}(k)_{2}$ Kac-Moody algebra,
and $\bar{\lambda} H_\text{LI}$ is the leading irrelevant term with coupling strength $\bar{\lambda} \propto (1/T_\text{K})^\Delta$.
Operators are labeled by quantum numbers $(Q,j_s,j_f)$ with charge $Q$, spin $j_s$, flavor $j_f$, respectively, in the 
$\mathrm{U}(1)$, $\mathrm{SU}(2)_{k}$, $\mathrm{SU}(k)_{2}$ sectors.
 
 
  
As in Eq.~\eqref{general_scaling1}, the impurity spin is identified~\cite{Barzykin98,Ludwig94} with a boundary operator $\vec{\psi}$ with scaling dimension $\Delta$,
\begin{align}\label{scaling}
S_{\mathrm{imp}}^{\alpha=x,y,z} \propto \frac{\psi^{\alpha}}{(T_\text{K})^\Delta} + \cdots.
\end{align}  
In the 1CK, $\vec{\psi}$ is the local spin density operator $\vec{J}$ with $\Delta = 1$ at the boundary.
In the $k$CK with $k \geq 2$, $\vec{\psi}$ is the $(Q=0,j_s=1, j_f=0)$ primary boundary operator $\vec{\phi}$ with $\Delta = 2 / (2+k)$.
\HS{Using Eq.~\eqref{general_scaling2}
and replacing $1/L$ by energy $E_i, \, E_j \sim E$  $(\ll T_\textrm{K})$, we find} 
\begin{align}\label{elementCFT}
\langle E_{i} | S_{\mathrm{imp}}^{\alpha=x,y,z} |E_{j} \rangle = O((\frac{E}{T_{\text{K}}})^{\Delta}),
\end{align}
\HS{which agrees with NRG results~\cite{supp}.}

Equation~\eqref{elementCFT} implies that 
each eigenstate $|E_i\rangle$ with $E_i \ll T_\textrm{K}$ is composed of 
a maximally entangled state $(|{\uparrow} \rangle \otimes |\phi_{i {\uparrow}} \rangle + |{\downarrow} \rangle \otimes |\phi_{i {\downarrow}} \rangle)/\sqrt{2}$ and small deviation $|\delta E_{i} \rangle$,
\begin{align}\label{stateT}
|E_{i} \rangle &= \frac{1}{\sqrt{2}} \Big( |{\uparrow} \rangle \otimes |\phi_{i{\uparrow}} \rangle + |{\downarrow} \rangle \otimes |\phi_{i{\downarrow}} \rangle \Big) + |\delta E_{i} \rangle.
\end{align}
$|\mu = {\uparrow}, {\downarrow} \rangle$ is the impurity spin state and
$|\phi_{i\mu} \rangle$'s are orthonormal states of the channels; the notation $|\phi_{i \mu} \rangle$ does not imply that the spin quantum number of $|\phi_{i \mu} \rangle$ is $\mu$~\cite{Singlet1}.
The impurity state is restored in the energy eigenstates $|E_i \rangle$.
Using Eq.~\eqref{elementCFT} and a Gram-Schmidt process~\cite{supp}, we derive \HS{Eq.~\eqref{stateT} and} find the scaling of  $\sqrt{\langle \delta E_{i} | \delta E_{i} \rangle} = O((E / T_{\text{K}})^{\Delta})$ and $\langle \delta E_{i} | E_{j} \rangle = O((E / T_{\text{K}})^{\Delta})$ in the restricted Hilbert space spanned by states of energy $E \sim E_{i} \sim E_j$.

\begin{figure}
\centerline{\includegraphics[width=.5\textwidth]{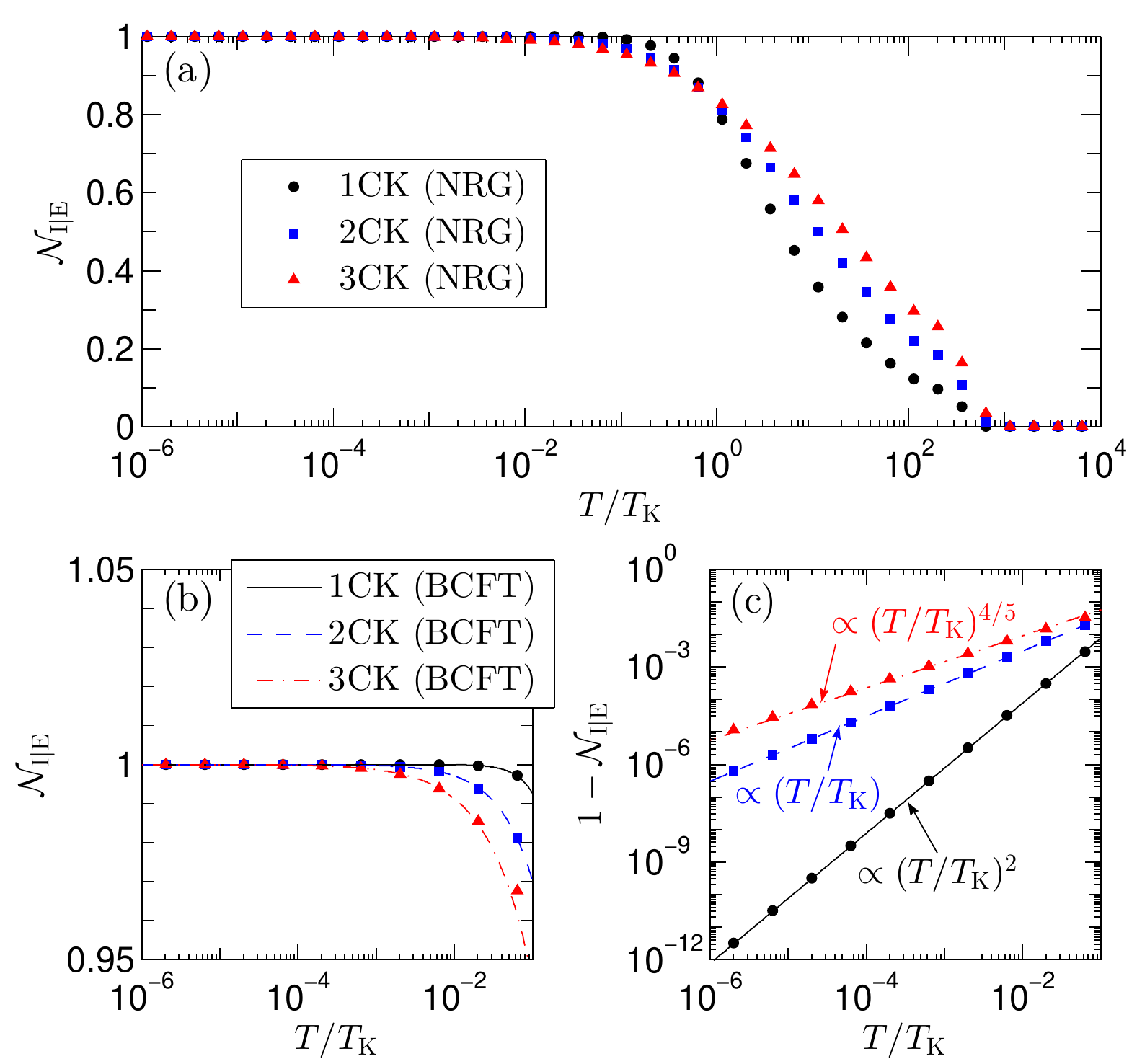}}
\caption{Temperature dependence of the negativity $\mathcal{N}_\textrm{I$|$E}$ between the impurity and screening channels in the isotropic $k$CK effects.
(a) NRG results~\cite{supp}, obtained by the method of Ref.~\cite{Shim18}. At $T=0$, $\mathcal{N}_\textrm{I$|$E} = 1$ independently of $k$. $\mathcal{N}_\textrm{I$|$E}$ rapidly decreases around $T_\textrm{K}$.
(b,c) NRG (dots) and BCFT (curves) results at $T \ll T_\text{K}$. $\mathcal{N}_\textrm{I$|$E}$ shows power-law scaling with $T/T_\text{K}$. The power-law exponents of the BCFT prediction in Eq.~\eqref{scalingT} agree with the NRG. (c) Log-log plot of $1- \mathcal{N}_\textrm{I$|$E} (T)$.
}
\label{MCKE_fig2}
\end{figure}

{\it Negativity at zero temperature.---}\HS{Equation~(\ref{stateT}) shows that each pure ground state is in the form $(|{\uparrow} \rangle \otimes |\phi_{g {\uparrow}} \rangle + |{\downarrow} \rangle \otimes |\phi_{g {\downarrow}} \rangle)/\sqrt{2}$, having maximal entanglement $\mathcal{N}_\textrm{I$|$E}=1$ between the impurity and the channels.} There can happen multiple $N_0$ degenerate ground states $|E_g =0 \rangle$'s in \HS{the $k$CK with $k \ge 2$}, e.g., when the channels satisfy the antiperiodic condition~\cite{Affleck91b}. Their thermal mixture $\rho (T=0) = \sum_{g=1}^{N_0} |E_g =0 \rangle\langle E_g=0| / N_0$ has the maximal entanglement, $\mathcal{N}_\textrm{I$|$E} = 1$, because $|\phi_{g \mu}\rangle$'s are mutually orthonormal.
This agrees with our NRG result in Fig.~\ref{MCKE_fig2}.

\HS{It is remarkable that the impurity is maximally entangled with the screening channels at zero temperature, independently of the channel number $k$.
In the 1CK, the perfect impurity screening originates from the maximal entanglement. In the $k$CK with $k \geq 2$, $\mathcal{N}_\textrm{I$|$E} = 1$ still happens, although the impurity is overscreened with $k$-dependent frustration.
This is in stark contrast with the impurity entropy~\cite{Sorensen07} $\zeta_\text{imp} = \ln g$ for the partition far outside the Kondo length [Fig.~\ref{MCKE_fig1}(b)] which is $k$ dependent; $g = 1$ for $k=1$ and $g = 2\cos[\pi/(2+k)]$ for $k \ge 2$.}

$\mathcal{N}_\textrm{I$|$E}$ and  $\zeta_\text{imp}$ reveal different \HS{but complementary} aspects of the Kondo effects.
$\mathcal{N}_\textrm{I$|$E}$ measures how the impurity is coherently screened or quantum correlation between the impurity and its screening cloud, while $\zeta_\text{imp}$ counts the effective ground-state degeneracy induced by the impurity~\cite{Affleck91a}. 
For example, $\mathcal{N}_\textrm{I$|$E} = 1$ shows that even at $k \to \infty$, the impurity is not free  but maximally correlated with the channels. Hence $\zeta_\text{imp} \to \ln 2$ at $k \to \infty$ does not imply a residual free moment. This is expected from the behavior of the impurity magnetization~\cite{Gan93,Gan94}. 

\HS{And, at $k = 2$, $\zeta_\text{imp} = \ln \sqrt{2}$ implies fractionalization of the impurity into two Majorana fermions, a free Majorana $\gamma_{i1}$ and another $\gamma_{i2}$ coupled with the channels.
Each ground state of the 2CK is a product state of a fermionic state by fusion of $\gamma_{i1}$ and a free Majorana of the channel, and another fusion state of $\gamma_{i2}$ and a Majorana of the channels~\cite{Zarand00}.
In terms of the bipartite basis states for the partition in Fig.~\ref{MCKE_fig1}(a), the product state is written as the maximal-entanglement form $(|{\uparrow} \rangle \otimes |\phi_{g {\uparrow}} \rangle + |{\downarrow} \rangle \otimes |\phi_{g {\downarrow}} \rangle)/\sqrt{2}$. Here the impurity states $|\uparrow \rangle$ and $|\downarrow\rangle$ are described by fusion of $\gamma_{i1}$ and $\gamma_{i2}$.
Hence, the fractionalization is reconciled with $\mathcal{N}_\textrm{I$|$E} = 1$. Similar maximal entanglement happens in a one-dimensional $p$-wave superconductor having free Majoranas~\cite{Park17}.
}

\HS{
{\it Universal thermal decay of negativity.---}
Putting $c_\alpha = 0$ and $\Delta_\alpha = \Delta$ of the $k$CK [see Eq.~\eqref{elementCFT}] into Eq.~\eqref{purestate_negativity3}, we find that the entanglement becomes weaker in the eigenstates of higher energy, $\mathcal{N}_\textrm{I$|$E} (| E_i \rangle) = 1 - O((E_i/T_{\text{K}})^{2 \Delta})$.
Then using Eq.~\eqref{purestate_negativity2}, we obtain the algebraic thermal decay of the entanglement in Eq.~\eqref{scalingT}.
The decay is confirmed by the NRG in Fig.~\ref{MCKE_fig2}, and also by direct calculation~\cite{supp} of \DHK{$\mathcal{N}_\textrm{I$|$E} + 1 = \Vert \rho^{\text{T}_\textrm{I}} \Vert = 2 - O((\frac{T}{T_\text{K}})^{2 \Delta})$} that equals the sum of the square roots of the eigenvalues of $(\rho^{\text{T}_\textrm{I}})^2$.
}

  

The power-law exponent $2 \Delta$ of the thermal decay of $\mathcal{N}_\textrm{I$|$E}$ is a universal fixed-point behavior. 
Considering the energy dependence $\langle \delta E_{i} | E_{j} \rangle = O((E / T_{\text{K}})^{\Delta})$ of the eigenstates in Eq.~\eqref{stateT}, it is nontrivial that the exponent of the entanglement decay is $2\Delta$ rather than $\Delta$; the contributions of $O((T / T_{\text{K}})^{\Delta})$ to $\mathcal{N}_\textrm{I$|$E}(\rho)$ exactly cancel each other. This relates with the fact that the entanglement is a nonlinear function of the density matrix.
The exponent ($\Delta$ versus $2 \Delta$) is determined by whether $c_\alpha = 0$ or not or equivalently by the ground-state expectation value of $\vec{S}_\textrm{imp}$; see Eqs.~\eqref{general_scaling1}-\eqref{purestate_negativity3}.
For example, we find~\cite{Ongoing} that the exponent of the algebraic thermal decay of $\mathcal{N}_\textrm{I$|$E}$ is $\Delta_z$ ($= \Delta_x$) in the Ising spin chain~\cite{Cardy89,Cardy91} with fixed boundary where $c_{x,z} \ne 0$ and $2 \Delta_y > \Delta_{x,z}$. 

\begin{figure}
\centerline{\includegraphics[width=.5\textwidth]{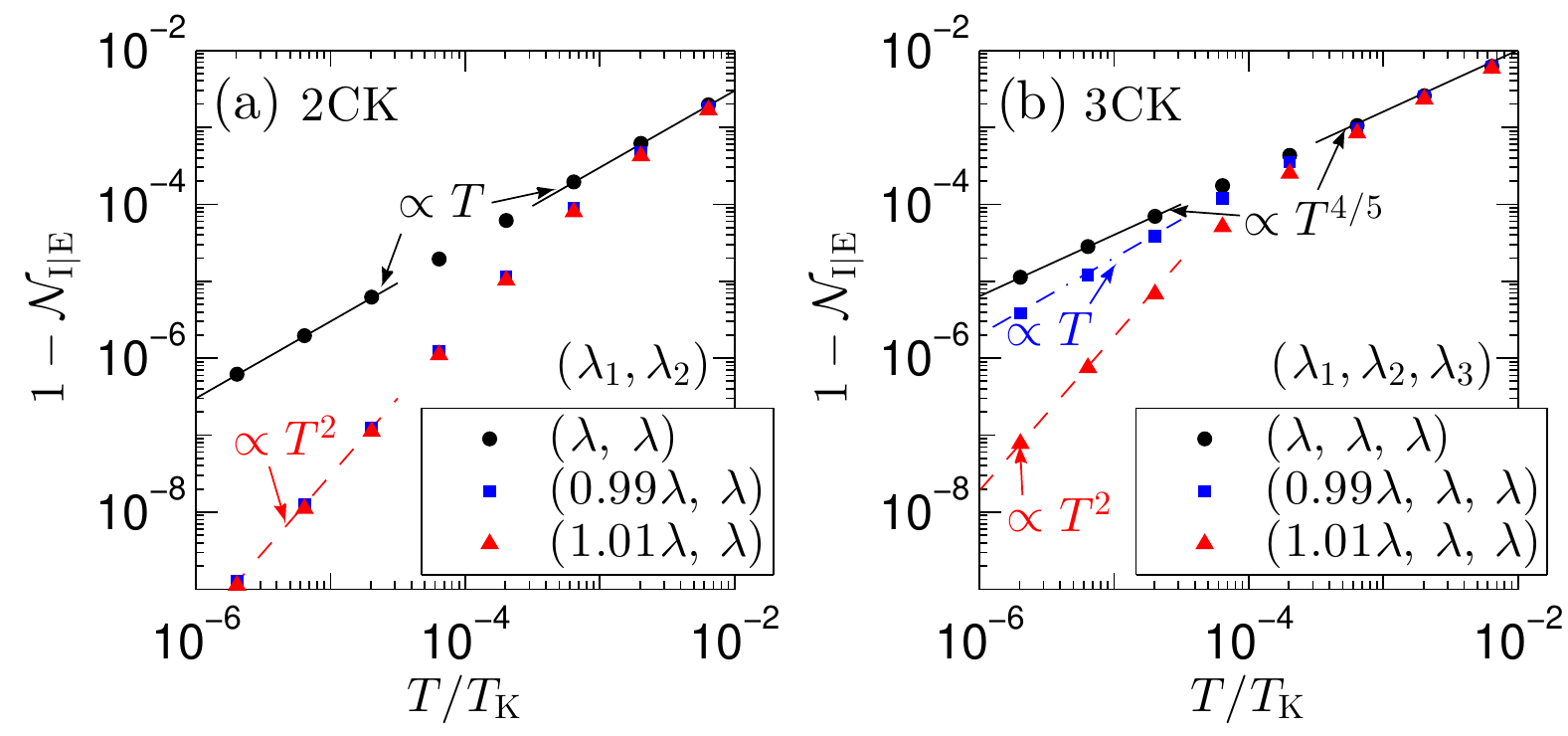}}
\caption{Temperature dependence of the negativity $\mathcal{N}_\textrm{I$|$E}$ in anisotropic (a) 2CK and (b) 3CK effects with the coupling strengths $\lambda_1$ and $\lambda_{i \ge 2} = \lambda$. The power-law exponent of $\mathcal{N}_\textrm{I$|$E}(T)$ changes, following crossover between different Kondo effects. The NRG results (points) agree with the BCFT prediction (lines). The isotropic case with $\lambda_1 = \lambda$ is shown for comparison.
(a) When $\lambda_1 = 0.99 \lambda$ or $1.01 \lambda$, the exponent changes from $2 \Delta = 2$ (the 1CK behavior) to $2 \Delta = 1$ (2CK) as $T$ increases, passing  the crossover temperature $T^*$. 
(b) When $\lambda_1 = 0.99 \lambda$, the exponent changes from $2 \Delta = 1$ (2CK) to $2 \Delta = 4/5$ (3CK).
When $\lambda_1 = 1.01 \lambda$, the exponent changes from $2 \Delta = 2$ (1CK) to $2 \Delta = 4/5$ (3CK).
}
\label{MCKE_fig3}
\end{figure}

{\it Negativity between fixed points.---}
We study the temperature dependence of the negativity $\mathcal{N}_\textrm{I$|$E}(T)$ in the channel anisotropic Kondo model, \HS{where thermal} crossover happens between different Kondo effects~\cite{Nozieres80,Vojta06}.
\HS{Figure~\ref{MCKE_fig3} shows that}
the power-law exponent of $\mathcal{N}_\textrm{I$|$E}(T)$ accordingly changes, following Eq.~\eqref{scalingT}.




We focus on the anisotropic 2CK model with the coupling strengths $\lambda_1 \ne \lambda_2$. Combining the finite-size bosonization method and our approach, 
we derive~\cite{supp}
\begin{align}\label{anisotropy_negativity}
1 - \mathcal{N}_\textrm{I$|$E} \propto
\left\{
\begin{array}{cc}
\frac{T}{T_\text{K}} & (T^* \ll T \ll T_\text{K}),
\\
\frac{\nu T^2}{T^*} & (T \ll T^*).
\end{array}
\right.
\end{align}
$T^*$ ($\propto \nu^2 (\lambda_2 - \lambda_1)^2 T_\text{K}$) is the crossover temperature between the 1CK at lower temperature and the 2CK at higher temperature. $\nu$ is the local density of  channel states at the impurity site.
Interestingly, the scaling behavior $\nu T^2 / T^*$ at $T \ll T^*$ is different from 
the known behavior of observables;  at $T \ll T^*$,  the decay of electron conductance in a setup follows $(T / T^*)^2$, while the magnetization follows $(T / T^*)^2 \sqrt{T^* T_\textrm{K}}$~\cite{Mitchell12}.
This scaling behavior of $\mathcal{N}_\textrm{I$|$E}$ is attributed~\cite{supp} to the scaling
 \begin{align}\label{anisotropy_spinmatrix}
\langle E_{i} | S_{\mathrm{imp}}^{\alpha} &|E_{j} \rangle =
\left\{
\begin{array}{cc}
O(\sqrt{\frac{E}{T_\text{K}}}) & (T^* \ll E \ll T_\text{K}) \\
O(\frac{\sqrt{\nu} E}{\sqrt{T^{*}}}) & (E \ll T^*)
\end{array}
\right.
\end{align}
with $E \sim E_i \sim E_j$.
This is confirmed by the NRG. 

{\it Conclusion.---}
We develop analytic computation of the entanglement $\mathcal{N}_\textrm{I$|$E}$ between the impurity and channels in the multichannel Kondo effects. $\mathcal{N}_\textrm{I$|$E}$ quantifies how the Kondo screening quantum coherently happens inside the screening length. Its thermal scaling is a universality of the fixed point and reflects non-Fermi liquids and fractionalization~\cite{Lopes20}.
$\mathcal{N}_\textrm{I$|$E}$ and the impurity entropy show different but complementary aspects of the Kondo effects.

Our findings have implications.
First, the direct relation in Eq.~\eqref{purestate_negativity1} between $\mathcal{N}_\textrm{I$|$E}$ and $\langle E_i | \vec{S}_\textrm{imp} | E_i \rangle$ for energy eigenstates means that the impurity spin screening originates from the entanglement $\mathcal{N}_\textrm{I$|$E}$ in general spin-1/2 impurities. It implies possibility of accessing $\mathcal{N}_\textrm{I$|$E}$ by experimentally detecting impurity magnetization at sufficiently low temperature. For example, in a quantum dot~\cite{Iftikhar15,Iftikhar18} showing multichannel charge Kondo effects, the excess charge of the dot, corresponding to $\langle E_i | \vec{S}_\textrm{imp} | E_i \rangle$, is detectable. One can measure $\mathcal{N}_\textrm{I$|$E}$ with changing a gate voltage applied to the dot, which corresponds to a magnetic field applied to $\vec{S}_\textrm{imp}$.

Second, the entanglement $\mathcal{N}_\textrm{I$|$E}$ will be useful for quantifying the spatial distribution of multichannel Kondo clouds.
The 1CK cloud was recently observed~\cite{Borzenets20} with changing the location at which the screening channel is weakly perturbed.
Similarly, we suggest~\cite{MCKC} to observe $k$CK clouds by monitoring change of $\mathcal{N}_\textrm{I$|$E}$ or  $\langle E_i | \vec{S}_\textrm{imp} | E_i \rangle$ with varying the perturbation position.

Third, the thermal scaling of $\mathcal{N}_\textrm{I$|$E}$ is a universal fixed-point property. This scaling behavior can be different from that of states, since $\mathcal{N}_\textrm{I$|$E}$ is a nonlinear function of the states. Interestingly, the thermal scaling of $\mathcal{N}_\textrm{I$|$E}$ is estimated from the scaling of $\mathcal{N}_\textrm{I$|$E}$ of energy eigenstates. This suggests to study entanglement in pure excited states in other problems, which has been less studied~\cite{Alba09,Calabrese15,Park17} than ground states.
   
\HS{Finally, our approach is applicable to study coherent coupling between the boundary and bulk in spin chains or in other spin-1/2 impurities described by BCFT. 
The behavior of $\mathcal{N}_\textrm{I$|$E}$ will depend on different universality classes of boundary phenomena of different boundary conditions.}

 
  


  




We thank Jan von Delft, Henrik Johannesson, Christophe Mora,  Fr\'{e}d\'{e}ric Pierre, June-Young M. Lee, and Seung-Sup B. Lee for insightful discussions. This work is supported by Korea NRF (SRC Center for Quantum Coherence in Condensed Matter, Grant No. 2016R1A5A1008184).

\clearpage

\onecolumngrid

\begin{center}
\textbf{\large Supplementary Material for \textquotedblleft Universal Scaling of Multi-channel Kondo Entanglement\textquotedblright}
\end{center}

\begin{center}
Author 1, Author 2, and Author 3

\textit{\small Department of Physics, Korea Advanced Institute of Science and Technology, Daejeon 34141, Korea}
\end{center}

\newcommand{\beginsup}{
        \setcounter{equation}{0}
        \renewcommand{\theequation}{S\arabic{equation}}
        \setcounter{table}{0}
        \renewcommand{\thetable}{S\arabic{table}}
        \setcounter{figure}{0}
        \renewcommand{\thefigure}{S\arabic{figure}}
     }
\beginsup

In this material, we prove that the maximum value of the negativity between the spin-1/2 impurity and the rest is $\mathcal{N}_\textrm{I$|$E} = 1$ in spin-1/2 quantum impurity problems.
It also contains NRG calculation details, NRG calculation of matrix elements of the impurity spin operator, and the validity of the approximate thermal density matrix. We provide the derivation of Eqs.~(3), (4), (6), (11), and (13), and direct computation of $\Vert \rho^{\text{T}_\textrm{I}} \Vert$ for the $k$CK effects.

\section{I. Maximum possible value of negativity}

We prove that the maximum possible value of the entanglement negativity $\mathcal{N}_\textrm{I$|$E}$ is 1 for the partition separating the spin-1/2 impurity from the rest in spin-1/2 quantum impurity problems.
Let $\mathcal{H}_{A}$, $\mathcal{H}_{B}$, and $\mathcal{H}$ be the Hilbert spaces of the impurity (a subsystem $A$), the other part of the total system (subsystem $B$), and the total system, respectively.
We calculate the negativity $\mathcal{N}_\textrm{I$|$E} = ||\rho^{\text{T}_A}|| - 1$.
The density matrix $\rho$ of the system is expressed as $\rho = \sum_{i=1}^{n} p_{i} |\Psi_{i} \rangle \langle \Psi_{i}|$ for some $|\Psi_{i} \rangle \in \mathcal{H}$ and non-negative real numbers $\{p_{i}\}_{i=1}^{n}$ with $\sum_{i=1}^{n} p_{i} = 1$.
Utilizing the fact that the dimension of $\mathcal{H}_{A}$ is 2, we express the states $|\Psi_{i} \rangle$ of the total system in the Schmidt decomposition form as $|\Psi_{i} \rangle = \sum_{j=1}^{2} a_{j}^{(i)} |\phi_{j}^{(i)} \rangle \otimes |\psi_{j}^{(i)} \rangle$ with  orthonormal-state sets $\{|\phi_{j}^{(i)}\rangle\}_{j=1}^{2} \subset \mathcal{H}_{A}$, $\{|\psi_{j}^{(i)} \rangle\}_{j=1}^{2} \subset \mathcal{H}_{B}$ and real numbers $\{a_{j}^{(i)}\}_{j=1}^{2}$ satisfying with $(a_{1}^{(i)})^{2} + (a_{2}^{(i)})^{2} = 1$.
Since the trace norm is a convex function, we obtain
\begin{align}
\Vert \rho^{\mathrm{T}_\textrm{I}} \Vert = \Bigg\Vert \Bigg(\sum_{i=1}^{n} p_{i} |\Psi_{i} \rangle \langle \Psi_{i}|\Bigg)^{\mathrm{T}_\textrm{I}} \Bigg\Vert \leq \sum_{i=1}^{n} p_{i} \Big\Vert \Big(|\Psi_{i} \rangle \langle \Psi_{i}|\Big)^{\mathrm{T}_\textrm{I}} \Big\Vert .
\end{align}
The partial transpose of each $|\Psi_{i} \rangle \langle \Psi_{i}|$ has the form of $(|\Psi_{i} \rangle \langle \Psi_{i}|)^{\mathrm{T}_\textrm{I}} = \sum_{j,k=1}^{2} a_{j}^{(i)} a_{k}^{(i)} (|\phi_{k}^{(i)} \rangle \otimes |\psi_{j}^{(i)} \rangle)(\langle \phi_{j}^{(i)}| \otimes \langle \psi_{k}^{(i)}|)$, and it has the eigenvalues of $(a_{1}^{(i)})^{2}$, $(a_{1}^{(i)})^{2}$, $a_{1}^{(i)} a_{2}^{(i)}$, and $-a_{1}^{(i)} a_{2}^{(i)}$.
These lead to
\begin{align}\label{purestate_tracenorm}
\Big\Vert \Big(|\Psi_{i} \rangle \langle \Psi_{i}|\Big)^{\mathrm{T}_\textrm{I}} \Big\Vert = (|a_{1}^{(i)}| + |a_{2}^{(i)}|)^{2} = 2 |a_{1}^{(i)}| |a_{2}^{(i)}| + 1 \leq 2, \qquad \because (a_{1}^{(i)})^{2} + (a_{2}^{(i)})^{2} = 1
\end{align}
implying $\Vert \rho^{\mathrm{T}_\textrm{I}} \Vert \leq \sum_{i} 2p_{i} = 2$.
Therefore, the negativity $\mathcal{N}_\textrm{I$|$E}$ must be less than or equal to $1$:
\begin{align}
\mathcal{N}_{\mathrm{I|E}} = \Vert \rho^{\mathrm{T}_\textrm{I}} \Vert - 1 \leq 1.
\end{align}

\section{II. Derivation of Eq.~(3)}
We derive Eq.~(3) and Eq.~(10) in the main text, based on the BCFT. 
We set the reduced Planck constant, Boltzmann constant, Fermi velocity as $\hbar=k_\text{B}=v =1$ henceforth.

\begin{figure}
\begin{center}
\begin{tikzpicture}
\filldraw [color=white,fill=red!5] (1,0) rectangle (6,3);
\draw [->] (3.5,-0.5) -- (3.5,3.1);
\draw [very thick] (1,0) -- (6,0);
\draw [->] (4.5,0) -- (6.2,0);
\draw (6.2,0) node[right]{$\mathrm{Re}(z)$};
\draw (3.5,3.1) node[above]{$\mathrm{Im}(z)$};
\draw (5,2.5) node{$z$};
\draw (3.3,0) node[below]{$0$};
\draw (-1,1.5) node[right]{$\xrightarrow{\displaystyle z = e^{\pi w / L}}$};
\filldraw [color=white,fill=red!5] (-7,0) rectangle (-2,2);
\draw [->] (-4.5,-0.5) -- (-4.5,3.1);
\draw [very thick] (-7,0) -- (-2,0);
\draw [very thick] (-7,2) -- (-2,2);
\draw [->] (-2.5,0) -- (-1.8,0);
\draw (-7,0) -- (-6.5,0);
\draw (-1.8,0) node[right]{$\mathrm{Re}(w)$};
\draw (-4.5,3.1) node[above]{$\mathrm{Im}(w)$};
\draw (-3,1) node{$w$};
\draw (-4.5,2.3) node[left]{$L$};
\draw (-4.7,0) node[below]{$0$};
\end{tikzpicture}
\end{center}
\caption{Conformal transformation from the infinite strip ($w$) to the half plane ($z$) through $z = e^{\pi w / L}$.
}\label{MCKE_supp_fig1}
\end{figure}
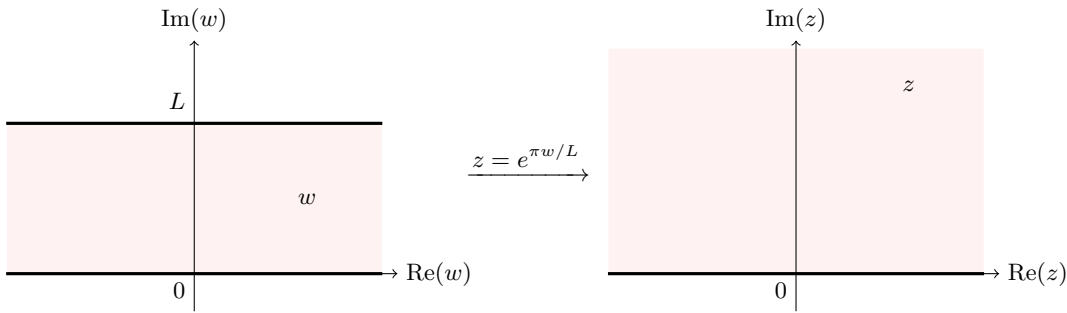

We consider a general spin-1/2 impurity described by BCFT Hamiltonian $H_\textrm{BCFT} = H_\text{FP} + \bar{\lambda} H_\text{LI}$ on the infinite strip with the width $L$ of holomorphic coordinate $w = \tau + i x$ ($\tau$ is the imaginary time, and $x$ is the spatial coordinate) [see the left panel of Fig.~\ref{MCKE_supp_fig1}].
We treat the conformally invariant fixed-point Hamiltonian $H_{\mathrm{FP}}$ as the bare Hamiltonian, and the leading irrelevant term  $\overline{\lambda} H_{\mathrm{LI}}$ as a perturbation.
The impurity spin $\vec{S}_{\mathrm{imp}}$ is identified as $S_{\mathrm{imp}}^{\alpha = x,y,z} = c_{\alpha} + d_{\alpha} \psi_{\alpha} + \cdots$ where $\psi_{\alpha}$ is a BCFT boundary operator with scaling dimension $\Delta_{\alpha}$ and $\cdots$ are operators of dimension $> \Delta_{\alpha}$.

We consider the $N$ lowest energy eigenstates $\{|E_{i} \rangle\}_{i=1}^{N}$ of the total Hamiltonian.
These states deviate from the eigenstates $\{|E_{i}^{0} \rangle\}$ of the bare Hamiltonian $H_{\mathrm{FP}}$ by the perturbation $\overline{\lambda} H_{\mathrm{LI}}$.

First, we show that the eigenstates $\{|E_{i}^{0} \rangle\}$ of $H_{\mathrm{FP}}$ satisfy
\begin{align}\label{bare matrix element scaling}
\langle E_{i}^{0} | \psi_{\alpha}(0) | E_{j}^{0} \rangle \propto \Bigg(\frac{\pi}{L}\Bigg)^{\Delta_{\alpha}}.
\end{align}
To obtain this $L$ dependence, it is convenient to use the conformal transformation in which  
the infinite strip of width $L$ is mapped into the upper half plane of  holomorphic coordinate $z$ by $z = e^{\pi w / L}$ [see Fig.~\ref{MCKE_supp_fig1}].
The boundary field $\psi_{\alpha}(\tau)$ in the strip is mapped to the boundary field $\psi_{\alpha}^{(H)}(e^{\pi \tau / L})$ where superscript $\,^{(H)}$ is introduced to distinguish the fields of the upper half-plane from those of the strip.
In the BCFT on the upper half plane, due to the boundary constraint, the set of modes $\mathcal{L}_{n}^{(H)} = \frac{1}{2 \pi i} \int_{C} z^{n+1} T(z) \, dz - \frac{1}{2 \pi i} \int_{C} \overline{z}^{n+1} \overline{T}(\overline{z}) \, d\overline{z}$ of the energy momentum tensors $T(z)$ and $\overline{T}(\overline{z})$ span a single Virasoro algebra, where the contour $C$ is a semi-circle in the upper half plane with the straight portion along the real line~\cite{Francesco97_supp,Recknagel13_supp}.
Then the boundary field satisfies the following relation~\cite{Recknagel13_supp}
\begin{align}\label{covariance of boundary field}
[\mathcal{L}_{n}^{(H)},\psi_{\alpha}^{(H)}(\tau)] = \Bigg(\tau^{n+1} \frac{d}{d\tau} + \Delta_{\alpha}(n+1)\tau^{n}\Bigg) \psi_{\alpha}^{(H)}(\tau) .
\end{align}
Equation~(\ref{covariance of boundary field}) gives the covariant relation $\psi_{\alpha}(\tau) = (\pi / L)^{\Delta_{\alpha}} \psi_{\alpha}^{(H)}(e^{\pi \tau / L})$ and we obtain
\begin{align}\label{bare matrix element}
\langle E_{i}^{0}| \psi_{\alpha}(0) | E_{j}^{0} \rangle = \Bigg(\frac{\pi}{L}\Bigg)^{\Delta_{\alpha}} \langle E_{i}^{0} | \psi_{\alpha}^{(H)}(1) | E_{j}^{0} \rangle .
\end{align}
Since $\langle E_{i}^{0} | \psi_{\alpha}^{(H)}(1) | E_{j}^{0} \rangle$ is independent of $L$ (namely, it is $O(1)$ due to the normalization of $|E_{i}^{0} \rangle$ and $|E_{j}^{0} \rangle$ and $\psi_{\alpha}^{(H)}$ being a boundary field on the upper half plane),
Eq.~(\ref{bare matrix element}) leads to Eq.~\eqref{bare matrix element scaling}.
Note that a result consistent with Eq.~\eqref{bare matrix element scaling} can be obtained by the approach of Ref.~\cite{Cardy86_supp,Affleck97_supp} based on the two-point and three-point correlation functions of $\vec{\psi}$.

Next, we show 
\begin{align}\label{matrix element scaling}
\langle E_{i} | \psi_{\alpha}(0) | E_{j} \rangle = O(L^{-\Delta_{\alpha}})
\end{align}
in terms of the low-energy eigenstates $|E_{i} \rangle$ of the total Hamiltonian $H_\textrm{BCFT}$. In the perturbation theory, the state $|E_{i} \rangle$ is written as $|E_{i} \rangle = |E_{i}^{0} \rangle + \sum_{n \neq i} |E_{n}^{0} \rangle \langle E_{n}^{0} | \overline{\lambda} H_{\mathrm{LI}} | E_{i}^{0} \rangle/(E_{i}^{0} - E_{n}^{0}) + O(\overline{\lambda}^{2})$.
Because $H_{\mathrm{LI}}$ is a scaling operator with dimension $1 + x$ ($x > 0$), the matrix elements with respect to the bare eigenstates satisfy $\langle E_{i}^{0} | \overline{\lambda} H_{\mathrm{LI}} | E_{j}^{0} \rangle \propto (\pi / L)^{1 + x}$ in the same way with Eq.~(\ref{bare matrix element scaling}). Combining it with $E_{i}^{0} - E_{n}^{0} \propto \pi / L$, we obtain $|E_{i} \rangle = |E_{i}^{0} \rangle + \sum_{n \neq i} O((\pi / L)^{x}) |E_{n}^{0} \rangle$. Thus, the leading $L$ dependence of $\langle E_{i} | \psi_{\alpha} |E_{j} \rangle$ comes from $\langle E_{i}^{0} |\psi_{\alpha}(0) | E_{j}^{0} \rangle$, and Eq.~\eqref{matrix element scaling} is obtained. 

Identification of the impurity spin $S^\alpha_\textrm{imp} = c_{\alpha} + d_{\alpha} \psi_{\alpha} + \cdots$, the normalization condition $\langle E_{i} | E_{j} \rangle = \delta_{ij}$ and Eq.~\eqref{matrix element scaling} lead to
\begin{align}\label{matrix element scaling2}
\langle E_{i} | S_{\mathrm{imp}}^{\alpha} | E_{j} \rangle = c_{\alpha} \delta_{ij} + d_{\alpha} O(L^{-\Delta_{\alpha}})
\end{align}
which is Eq.~(3) of the main text.

Then Eq.~(10) in the main text is derived straightforwardly from Eq.~\eqref{matrix element scaling2}.
In the $k$CK model, the impurity spin $\vec{S}_{\mathrm{imp}}$ is identified as in Eq.~(8) of the main text; $c_{\alpha} = 0$, $d_{\alpha} = 1 / T_\textrm{K}^{\Delta}$, and $\Delta_{\alpha} = \Delta$ for $\alpha = x,y,z$. Replacing $1/L$ by $E \sim E_{i} \sim E_{j}$, we obtain Eq.~(10).

\section{III. Derivation of Eq.~(4)}
We derive Eq.~(4) in the main text. The state $|E_{i} \rangle$ has a general form $|E_{i} \rangle = |{\uparrow} \rangle |\widetilde{\psi}_{i {\uparrow}} \rangle + |{\downarrow} \rangle |\widetilde{\psi}_{i{\downarrow}} \rangle$, where $|\mu = {\uparrow},{\downarrow} \rangle$ are the impurity spin eigenstates of $S_{\mathrm{imp}}^{z}$, and $|\widetilde{\psi}_{i \mu} \rangle$'s are unnormalized channel states. By the normalization condition of $|E_{i} \rangle$ and the action of $S_{\mathrm{imp}}^{z}$ and $S_{\mathrm{imp}}^{-} = S_{\mathrm{imp}}^{x} - i S_{\mathrm{imp}}^{y}$ on $|E_{i} \rangle$, we find
\begin{equation}
\begin{aligned}
\langle \widetilde{\psi}_{i {\uparrow}} | \widetilde{\psi}_{i {\uparrow}} \rangle + \langle \widetilde{\psi}_{i {\downarrow}} | \widetilde{\psi}_{i {\downarrow}} \rangle &= 1 \\
\langle \widetilde{\psi}_{i {\uparrow}} | \widetilde{\psi}_{i {\uparrow}} \rangle - \langle \widetilde{\psi}_{i {\downarrow}} | \widetilde{\psi}_{i {\downarrow}} \rangle &= 2 \langle E_{i} | S_{\mathrm{imp}}^{z} | E_{i} \rangle \\
\langle \widetilde{\psi}_{i {\downarrow}} | \widetilde{\psi}_{i {\uparrow}} \rangle &= \langle E_{i} | S_{\mathrm{imp}}^{-} | E_{i} \rangle.
\end{aligned}
\end{equation}
Notice that a state $|E_i \rangle$ is separable, $\mathcal{N}_{\mathrm{I|E}}(|E_{i} \rangle) = 0$, when $\langle E_{i} |S_{\mathrm{imp}}^{z} |E_{i} \rangle = -1/2$. In this case, $|E_{i} \rangle = |{\downarrow} \rangle |\widetilde{\psi}_{i\downarrow} \rangle$ with $\langle \widetilde{\psi}_{i \downarrow} | \widetilde{\psi}_{i \downarrow} \rangle = 1$. Below, we consider the cases of $\langle E_{i} | S_{\mathrm{imp}}^{z} |E_{i} \rangle \neq - 1/2$.

Introducing orthonormal channel states $|\phi_{i {\uparrow}} \rangle \equiv \frac{|\widetilde{\psi}_{i {\uparrow}} \rangle}{\sqrt{\langle \widetilde{\psi}_{i {\uparrow}} | \widetilde{\psi}_{i {\uparrow}} \rangle}}$ and $|\phi_{i {\downarrow}} \rangle \equiv \frac{(1 - |\phi_{i {\uparrow}} \rangle \langle \phi_{i {\uparrow}}|)|\widetilde{\psi}_{i {\downarrow}} \rangle}{\sqrt{\langle \widetilde{\psi}_{i {\downarrow}} | (1 - |\phi_{i {\uparrow}} \rangle \langle \phi_{i {\uparrow}}|) | \widetilde{\psi}_{i {\downarrow}} \rangle}}$, we rewrite $|E_{i} \rangle$ as
\begin{align}\label{standardform}
|E_{i} \rangle = a |{\uparrow} \rangle |\phi_{i {\uparrow}} \rangle + b |{\downarrow} \rangle |\phi_{i{\uparrow}} \rangle + c |{\downarrow} \rangle |\phi_{i {\downarrow}} \rangle
\end{align}
where $a = \sqrt{\frac{1 + 2 \langle E_{i} | S_{\mathrm{imp}}^{z} | E_{i} \rangle}{2}}$, $b = \sqrt{\frac{2}{1 + 2 \langle E_{i} | S_{\mathrm{imp}}^{z} | E_{i} \rangle}} \langle E_{i} | S_{\mathrm{imp}}^{-} | E_{i} \rangle^{*}$, and $c = \sqrt{\frac{1 - 4 \langle E_{i} | S_{\mathrm{imp}}^{z} | E_{i} \rangle^{2} - 4 |\langle E_{i} | S_{\mathrm{imp}}^{-} | E_{i} \rangle|^{2}}{2(1 + 2 \langle E_{i} | S_{\mathrm{imp}}^{z} | E_{i} \rangle)}}$. By Schmidt decomposition, there exist orthonormal impurity states $\{|{\uparrow}' \rangle, |{\downarrow}' \rangle\}$ and orthonormal channel states $\{|\phi_{{\uparrow}}' \rangle, |\phi_{{\downarrow}}' \rangle\}$ such that
\begin{align}
|E_{i} \rangle = \lambda_{1} |{\uparrow}' \rangle |\phi_{{\uparrow}}' \rangle + \lambda_{2} |{\downarrow}' \rangle |\phi_{{\downarrow}}' \rangle.
\end{align}
Here, $\lambda_{1}$ and $\lambda_{2}$ are the singular values of the matrix $\begin{pmatrix} a & 0 \\
b & c \end{pmatrix}$, which are explicitly $\sqrt{\frac{a^{2} + |b|^{2} + c^{2}}{2} \pm \sqrt{(\frac{a^{2} + |b|^{2} + c^{2}}{2})^{2} - a^{2} c^{2}}}$. The negativity $\mathcal{N}_{\mathrm{I|E}}(|E_{i} \rangle)$ is determined by the product of two coefficients $\lambda_{1}$, $\lambda_{2}$ as shown in Eq.~(\ref{purestate_tracenorm}),
\begin{align}
\mathcal{N}_{\mathrm{I|E}}(|E_{i} \rangle) = 2 \lambda_{1} \lambda_{2} = 2 ac = \sqrt{1 - 4 \langle E_{i} | S_{\mathrm{imp}}^{z} | E_{i} \rangle^{2} - 4 |\langle E_{i} | S_{\mathrm{imp}}^{-} | E_{i} \rangle|^{2}}.
\end{align}
Since $|\langle E_{i} | S_{\mathrm{imp}}^{-} |E_{i} \rangle|^{2} = \langle E_{i} | S_{\mathrm{imp}}^{x} | E_{i} \rangle^{2} + \langle E_{i} | S_{\mathrm{imp}}^{y} | E_{i} \rangle^{2}$, we finally obtain the Eq. (4) of the main text,
\begin{align}\label{pure_negativity}
\mathcal{N}_{\mathrm{I|E}}(|E_{i} \rangle) = \sqrt{1 - 4 \langle E_{i} | S_{\mathrm{imp}}^{x} | E_{i} \rangle^{2} - 4 \langle E_{i} | S_{\mathrm{imp}}^{y} | E_{i} \rangle^{2} - 4 \langle E_{i} | S_{\mathrm{imp}}^{z} | E_{i} \rangle^{2}} = \sqrt{1 - 4 \langle E_{i} | \vec{S}_{\mathrm{imp}} | E_{i} \rangle^{2}}
\end{align}
where $\vec{S}_{\mathrm{imp}} = S_{\mathrm{imp}}^{x} \mathbf{e}_{x} + S_{\mathrm{imp}}^{y} \mathbf{e}_{y} + S_{\mathrm{imp}}^{z} \mathbf{e}_{z}$ is the spin-1/2 angular momentum. 
Equation~\eqref{pure_negativity} includes the result $\mathcal{N}_{\mathrm{I|E}}(|E_{i} \rangle) = 0$ of the case of $\langle E_{i}| S_{\mathrm{imp}}^{z} |E_{i} \rangle = -1/2$, so it holds for any state $|E_{i} \rangle$.

\section{IV. Validity of the approximate thermal density matrix}

In the main text, we construct an approximate thermal density matrix $\rho$ at temperature $T$ in an effective Hilbert space which is constituted by the energy eigenstates having energy $E \sim T$.
We below discuss the validity of the approximate density matrix.

The validity of this approximate density matrix has been rigorously justified in the NRG.
The corresponding approximation used in the NRG approach is called the single-shell approximation~\cite{Weichselbaum07_supp,Weichselbaum12_supp}, which we briefly explain below.
In the NRG, 
the complete basis $\cup_n \{|E_{ni}^D\rangle\otimes|\vec{s}_n\rangle\}_{i=1}^N$ of the Wilson chains is constructed~\cite{Anders05_supp} by the discarded states $\{|E_{ni}^D\rangle\}_{i=1}^N$ at every NRG iteration step $n$.
Here $|E_{ni}^D\rangle$ is the $i$th discarded state at the $n$th NRG iteration step having energy $E_{ni}^D \sim O(\Lambda^{-n/2})$,
$\Lambda > 1$ is a numerical discretization parameter to ensure the energy scale separation between different NRG iteration steps,
$N$ is the number of the discarded states at the $n$th NRG iteration step,
$|\vec{s}_n\rangle=|s_{n+1}\rangle\otimes\cdots\otimes|s_{N_\mathrm{WC}}\rangle$ is environment states describing from the ($n+1$)th site ($|s_{n+1}\rangle$) to the $N_\mathrm{WC}$th site ($|s_{N_\mathrm{WC}}\rangle$) of the Wilson chains, 
and $N_\mathrm{WC}$ is the number of the sites of each Wilson chain. Using the complete basis states, the NRG full density matrix at temperature $T$ is written as
\begin{align}
\rho_\mathrm{FDM}(T) = \sum_{n}\sum_{i=1}^N\sum_{\vec{s}_n} \frac{e^{-E_{ni}^D/T}}{Z}|E_{ni}^D\rangle\langle E_{ni}^D| \otimes |\vec{s}_n\rangle\langle \vec{s}_n|,
\end{align}
where $Z = \sum_{n}\sum_{i=1}^N e^{-E_{ni}^D/T} d_s^{N_\mathrm{WC}-n}$ and
$d_s$ is the degree of freedom of each environment state $|s_{n+1,\cdots,N_\mathrm{WC}}\rangle$.
The full density matrix is decomposed into the density matrices $\rho_n$ of the discarded states $|E_{ni}^D\rangle$ at the $n$th NRG iteration step,
\begin{align}
\rho_\mathrm{FDM} = \sum_{n} \frac{Z_{n} d_s^{N_\mathrm{WC}-n}}{Z}\rho_n \otimes \Big(\frac{1}{d_s^{N_\mathrm{WC}-n}} \sum_{\vec{s}_n}|\vec{s}_n\rangle\langle \vec{s}_n|\Big),
\end{align}
where $\rho_n = \sum_{i=1}^N e^{-E_{ni}^D/T}/Z_n |E_{ni}^D\rangle\langle E_{ni}^D|$ and $Z_n = \sum_{i=1}^N e^{-E_{ni}^D/T}$.
Note that both the density matrices $\rho_\mathrm{FDM}$ and $\rho_n$ are normalized as $\mathrm{tr}(\rho_\mathrm{FDM}) = \mathrm{tr}(\rho_n) = 1$.
In the single-shell approximation, the full density matrix $\rho_\mathrm{FDM}$ at temperature $T$ is approximated by the density matrix $\rho_{N_T}$ of the NRG iteration step $n=N_T$ at which the discarded states have the energy $O(\Lambda^{-N_T /2}) \sim T$.
This approximation has been justified by the fact~\cite{Weichselbaum12_supp} that the distribution $Z_n d_s^{N_\mathrm{WC}-n} / Z$ is peaked at $n = N_T$ and the discarded states $|E_{N_T,i}^D\rangle$ at the $N_T$th step dominate the full density matrix $\rho_\mathrm{FDM}$.
 
The approximate density matrix in the main text, formed by the BCFT or bosonization method with finite size $L \sim 1/T$, is equivalent with the density matrix $\rho_{N_T}$ obtained with the single-shell approximation  where  $O(\Lambda^{-N_T /2}) \sim T$.
It is because the energy spectrum and degeneracy of a finite-size system obtained by the NRG are consistent with those obtained by the BCFT~\cite{Affleck91b_supp,Affleck92_supp} or the bosonization method~\cite{Zarand00_supp}.
Therefore, the approximate thermal density matrix $\rho$ at temperature $T$, used in the main text, can be constructed by the low energy states $\{|E_{i} \rangle\}_{i=1}^{N}$ of the BCFT or bosonization Hamiltonian of finite size $L \sim 1/T$. Note that the low-energy states have energy $E_{i} \sim T$. The approximate density matrix $\rho$ has the following form
\begin{align}\label{densitymatrixform}
\rho = \sum_{i=1}^{N} \frac{e^{- E_{i} / T}}{Z} |E_{i} \rangle \langle E_{i}| = \sum_{i=1}^{N} w_{i} |E_{i} \rangle \langle E_{i}|
\end{align}
where $Z = \sum_{i=1}^{N} e^{-E_{i}/T}$ is the partition function in the effective Hilbert space formed by the low-energy states $\{|E_{i} \rangle\}_{i=1}^{N}$ and $w_{i} = e^{-E_{i}/T} / Z$. Here $N$ is not too large, as $E_N \sim T$.
 
\section{V. Derivation of Eq.~(6)}
We derive Eq.~(6) of the main text. For the purpose, we first represent the energy eigenstates $|E_{i} \rangle$'s in the bipartite orthonormal basis, and then we calculate $\mathcal{N}_{\mathrm{I|E}}(\rho)$.
 
\subsection{State representation in the bipartite orthonormal basis}
We consider a restricted Hilbert space spanned by states of energy $\sim E$. We start from a general form of $|E_{i} \rangle$, $|E_{i} \rangle = |{\uparrow} \rangle  |\widetilde{\psi}_{i{\uparrow}} \rangle + |{\downarrow} \rangle |\widetilde{\psi}_{i{\downarrow}} \rangle$, where $|\mu = {\uparrow},{\downarrow} \rangle$'s are the impurity spin eigenstates of $S_{\mathrm{imp}}^{z}$, and $|\widetilde{\psi}_{i{\mu}} \rangle$'s are unnormalized channel states.
The normalization $\langle E_{i} | E_{j} \rangle = \delta_{ij}$ and the action of the impurity spin $S_{\mathrm{imp}}^{z}$ and $S_{\mathrm{imp}}^{-} = S_{\mathrm{imp}}^{x} - i S_{\mathrm{imp}}^{y}$ on $|E_{i} \rangle$ yield
\begin{align}
\langle \widetilde{\psi}_{i {\uparrow}} | \widetilde{\psi}_{j {\uparrow}} \rangle + \langle \widetilde{\psi}_{i{\downarrow}} | \widetilde{\psi}_{j{\downarrow}} \rangle &= \delta_{ij}, &&\qquad \forall i,j \nonumber \\
\langle \widetilde{\psi}_{i {\uparrow}} | \widetilde{\psi}_{j {\uparrow}} \rangle - \langle \widetilde{\psi}_{i{\downarrow}} | \widetilde{\psi}_{j{\downarrow}} \rangle &= 2 \langle E_{i} | S_{\mathrm{imp}}^{z} | E_{j} \rangle, &&\qquad \forall i,j \label{scaling_supp5-1} \\
\langle \widetilde{\psi}_{i {\downarrow}} | \widetilde{\psi}_{j {\uparrow}} \rangle &= \langle E_{i} | S_{\mathrm{imp}}^{-} | E_{j} \rangle &&\qquad \forall i,j. \nonumber
\end{align}
Replacing $1/L$ by $E \sim E_{i} \sim E_{j}$ in Eq.~\eqref{matrix element scaling2}, the scaling behavior of the impurity spin becomes
\begin{align}\label{energyscaling}
\langle E_{i} | S_{\mathrm{imp}}^{\alpha} | E_{j} \rangle = c_{\alpha} \delta_{ij} + d_{\alpha} O(E^{\Delta_{\alpha}}) \quad \quad \forall \alpha = x,y,z.
\end{align}
Below we derive Eq.~(6) of the main text by applying a specific Gram-Schmidt orthogonalization process to $|\widetilde{\psi}_{i \mu} \rangle$'s, which is for the case of $\sqrt{c_x^2 + c_y^2 + c_z^2 }< 1 /2$ where the ground states have the entanglement between the impurity and the environment [see Eq.~(5) of the main text]. Equation (6) is derived also for the other case of $\sqrt{c_x^2 + c_y^2 + c_z^2 } = 1 /2$ by applying another Gram-Schmidt orthogonalization process (not shown).
 

By applying the Gram-Schmidt process to $|\widetilde{\psi}_{i \mu} \rangle$'s, we introduce the following states satisfying $\langle \phi_{i\mu} | \phi_{j\eta} \rangle = \delta_{ij} \delta_{\mu \eta}$
\begin{align}
|\phi_{1 {\uparrow}} \rangle &\equiv \frac{|\widetilde{\psi}_{1 {\uparrow}} \rangle}{\sqrt{\langle \widetilde{\psi}_{1 {\uparrow}} |\widetilde{\psi}_{1 {\uparrow}} \rangle}}, \label{Gram_Schmidt5-1} \\
|\phi_{1 {\downarrow}} \rangle &\equiv \frac{(1 - |\phi_{1 {\uparrow}} \rangle \langle \phi_{1 {\uparrow}}|)|\widetilde{\psi}_{1 {\downarrow}} \rangle}{\sqrt{\langle \widetilde{\psi}_{1 {\downarrow}}| (1 - |\phi_{1 {\uparrow}} \rangle \langle \phi_{1 {\uparrow}}|)|\widetilde{\psi}_{1 {\downarrow}} \rangle}}, \label{Gram_Schmidt5-2} \\
|\phi_{2 {\uparrow}} \rangle &\equiv \frac{(1 - |\phi_{1 {\uparrow}} \rangle \langle \phi_{1 {\uparrow}}| - |\phi_{1 {\downarrow}} \rangle \langle \phi_{1 {\downarrow}}|)|\widetilde{\psi}_{2 {\uparrow}} \rangle}{\sqrt{\langle \widetilde{\psi}_{2 {\uparrow}}| (1 - |\phi_{1 {\uparrow}} \rangle \langle \phi_{1 {\uparrow}}| - |\phi_{1 {\downarrow}} \rangle \langle \phi_{1 {\downarrow}}|)|\widetilde{\psi}_{2 {\uparrow}} \rangle}}, \label{Gram_Schmidt5-3} \\
|\phi_{2 {\downarrow}} \rangle &\equiv \frac{(1 - |\phi_{1 {\uparrow}} \rangle \langle \phi_{1 {\uparrow}}| - |\phi_{1 {\downarrow}} \rangle \langle \phi_{1 {\downarrow}}| - |\phi_{2 {\uparrow}} \rangle \langle \phi_{2 {\uparrow}}|)|\widetilde{\psi}_{2 {\downarrow}} \rangle}{\sqrt{\langle \widetilde{\psi}_{2 {\downarrow}}| (1 - |\phi_{1 {\uparrow}} \rangle \langle \phi_{1 {\uparrow}}| - |\phi_{1 {\downarrow}} \rangle \langle \phi_{1 {\downarrow}}| - |\phi_{2 {\uparrow}} \rangle \langle \phi_{2 {\uparrow}}|)|\widetilde{\psi}_{2 {\downarrow}} \rangle}}. \label{Gram_Schmidt5-4} \\
\vdots  & \nonumber
\end{align}
We then write $|\widetilde{\psi}_{i \mu} \rangle $ in terms of orthonormal states $|\phi_{i \mu} \rangle$'s, using Eqs.~\eqref{scaling_supp5-1}-\eqref{Gram_Schmidt5-4}, as 
\begin{align}
|\widetilde{\psi}_{1 \uparrow} \rangle & = \sqrt{\langle \widetilde{\psi}_{1 \uparrow} | \widetilde{\psi}_{1 \uparrow} \rangle} |\phi_{1 \uparrow} \rangle = \sqrt{\frac{1 + 2 \langle E_{1} | S_{\mathrm{imp}}^{z} | E_{1} \rangle}{2}} |\phi_{1 \uparrow} \rangle \\
|\widetilde{\psi}_{1 \downarrow} \rangle & = \sqrt{\frac{2}{1 + 2 \langle E_{1} | S_{\mathrm{imp}}^{z} | E_{1} \rangle}} \langle E_{1} | S_{\mathrm{imp}}^{-} | E_{1} \rangle^{*} |\phi_{1 \uparrow} \rangle + \sqrt{\frac{1 - 4 \langle E_{1} | S_{\mathrm{imp}}^{z} | E_{1} \rangle^{2} - 4 |\langle E_{1} | S_{\mathrm{imp}}^{-} | E_{1} \rangle|^{2}}{2(1 + 2 \langle E_{1} | S_{\mathrm{imp}}^{z} | E_{1} \rangle)}} |\phi_{1 \downarrow} \rangle \\
|\widetilde{\psi}_{2 \uparrow} \rangle & = \Bigg(\sqrt{\frac{1 + 2 \langle E_{2} | S_{\mathrm{imp}}^{z} | E_{2} \rangle}{2}} + O(E^{\mathrm{min}\{2\Delta_{x},2\Delta_{y},2\Delta_{z}\}}) \Bigg) |\phi_{2 \uparrow} \rangle + O(E^{\Delta_{z}}) |\phi_{1\uparrow} \rangle + \Big(O(E^{\Delta_{-}})+ O(E^{\Delta_{z}})\Big) |\phi_{1 \downarrow} \rangle  \label{Gram_Schmidt_2up} \\
|\widetilde{\psi}_{2 \downarrow} \rangle &= 
\Bigg(\sqrt{\frac{2}{1 + 2 \langle E_{2} | S_{\mathrm{imp}}^{z} | E_{2} \rangle}} \langle E_{2} | S_{\mathrm{imp}}^{-} | E_{2} \rangle^{*} +  O(E^{\mathrm{min}\{2\Delta_{x},2\Delta_{y},2\Delta_{z}\}}) \Bigg) |\phi_{2 \uparrow} \rangle \nonumber \\
& + \Bigg(\sqrt{\frac{1 - 4 \langle E_{2} | S_{\mathrm{imp}}^{z} | E_{2} \rangle^{2} - 4 |\langle E_{2} | S_{\mathrm{imp}}^{-} | E_{2} \rangle|^{2}}{2(1 + 2 \langle E_{2} | S_{\mathrm{imp}}^{z} | E_{2} \rangle)}} +  O(E^{\mathrm{min}\{2\Delta_{x},2\Delta_{y},2\Delta_{z}\}}) \Bigg) |\phi_{2 \downarrow} \rangle \nonumber \\
& + O(E^{\Delta_{-}}) |\phi_{1\uparrow} \rangle + \Big(O(E^{\Delta_{z}})+ O(E^{\Delta_{-}})\Big) |\phi_{1 \downarrow} \rangle  \label{Gram_Schmidt_2down} \\
\vdots &   \nonumber \\
|\widetilde{\psi}_{i\uparrow} \rangle & = \Bigg(\sqrt{\frac{1 + 2 \langle E_{i} | S_{\mathrm{imp}}^{z} | E_{i} \rangle}{2}} + O(E^{\mathrm{min}\{2\Delta_{x},2\Delta_{y},2\Delta_{z}\}})\Bigg) |\phi_{i \uparrow} \rangle + \sum_{m < i} \sum_{\eta = \uparrow,\downarrow} \Big(O(E^{\Delta_{z}}) + O(E^{\Delta_{-}})\Big) |\phi_{m \eta} \rangle \label{Gram_Schmidt_up}  \\
|\widetilde{\psi}_{i\downarrow} \rangle &= \Bigg(\sqrt{\frac{2}{1 + 2 \langle E_{i} | S_{\mathrm{imp}}^{z} | E_{i} \rangle}} \langle E_{i} | S_{\mathrm{imp}}^{-} | E_{i} \rangle^{*} + O(E^{\mathrm{min}\{2\Delta_{x},2\Delta_{y},2\Delta_{z}\}})\Bigg) |\phi_{i \uparrow} \rangle \nonumber \\
& + \Bigg(\sqrt{\frac{1 - 4 \langle E_{i} | S_{\mathrm{imp}}^{z} | E_{i} \rangle^{2} - 4 |\langle E_{i} | S_{\mathrm{imp}}^{-} | E_{i} \rangle|^{2}}{2(1 + 2 \langle E_{i} | S_{\mathrm{imp}}^{z} | E_{i} \rangle)}} + O(E^{\mathrm{min}\{2\Delta_{x},2\Delta_{y},2\Delta_{z}\}})\Bigg) |\phi_{i \downarrow} \rangle \nonumber \\
& + \sum_{m < i} \sum_{\eta = \uparrow,\downarrow} \Big(O(E^{\Delta_{z}}) + O(E^{\Delta_{-}})\Big) |\phi_{m \eta} \rangle.\label{Gram_Schmidt_down}
\end{align}
Here, $\Delta_-$ is the scaling dimension of $S^-_\textrm{imp}$, and $\mathrm{min}\{2\Delta_{x},2\Delta_{y},2\Delta_{z}\}$ is the mininum among $2 \Delta_x$, $2 \Delta_y$, and $2 \Delta_z$; $\Delta_-$ is determined by $\mathrm{min} \{ \Delta_x, \Delta_y \}$.
Using the above expressions of $|\widetilde{\psi}_{i\uparrow} \rangle$ and $|\widetilde{\psi}_{i\downarrow} \rangle$, we write $|E_{i} \rangle = |{\uparrow} \rangle  |\widetilde{\psi}_{i{\uparrow}} \rangle + |{\downarrow} \rangle |\widetilde{\psi}_{i{\downarrow}} \rangle$ in the decomposition of
\begin{align}\label{Decomposition1}
|E_{i} \rangle  = |{\uparrow} \rangle |\widetilde{\psi}_{i \uparrow} \rangle + |{\downarrow} \rangle |\widetilde{\psi}_{i \downarrow} \rangle = |e_{i} \rangle + |\delta e_{i} \rangle
\end{align}
where $|e_{i} \rangle$ is the projection of $|E_{i} \rangle$ onto the subspace $\{|{\uparrow} \rangle |\phi_{i \uparrow} \rangle,|{\uparrow} \rangle |\phi_{i \downarrow} \rangle,|{\downarrow} \rangle |\phi_{i \uparrow} \rangle,|{\downarrow} \rangle |\phi_{i \downarrow} \rangle\}$ and $|\delta e_{i} \rangle$ is the remainder,
\begin{equation}\label{Decomposition2}
\begin{aligned}
|e_{i} \rangle &= \Bigg(\sqrt{\frac{1 + 2 \langle E_{i} | S_{\mathrm{imp}}^{z} | E_{i} \rangle}{2}} + O(E^{\mathrm{min}\{2\Delta_{x},2\Delta_{y},2\Delta_{z}\}})\Bigg) |{\uparrow} \rangle |\phi_{i \uparrow} \rangle \\
& + \Bigg(\sqrt{\frac{2}{1 + 2 \langle E_{i} | S_{\mathrm{imp}}^{z} | E_{i} \rangle}} \langle E_{i} | S_{\mathrm{imp}}^{-} | E_{i} \rangle^{*} + O(E^{\mathrm{min}\{2\Delta_{x},2\Delta_{y},2\Delta_{z}\}})\Bigg) |{\downarrow} \rangle |\phi_{i \uparrow} \rangle \\
& + \Bigg(\sqrt{\frac{1 - 4 \langle E_{i} | S_{\mathrm{imp}}^{z} | E_{i} \rangle^{2} - 4 |\langle E_{i} | S_{\mathrm{imp}}^{-} | E_{i} \rangle|^{2}}{2(1 + 2 \langle E_{i} | S_{\mathrm{imp}}^{z} | E_{i} \rangle)}} + O(E^{\mathrm{min}\{2\Delta_{x},2\Delta_{y},2\Delta_{z}\}})\Bigg) |{\downarrow} \rangle |\phi_{i \downarrow} \rangle,
\end{aligned}
\end{equation}
\begin{align}\label{Decomposition3}
|\delta e_{i} \rangle = \sum_{m < i} \sum_{\eta = \uparrow,\downarrow} \Big(O(E^{\Delta_{z}}) + O(E^{\Delta_{-}})\Big) |{\uparrow} \rangle |\phi_{m \eta} \rangle + \sum_{m < i} \sum_{\eta = \uparrow,\downarrow} \Big(O(E^{\Delta_{z}}) + O(E^{\Delta_{-}})\Big) |{\downarrow} \rangle |\phi_{m \eta} \rangle.
\end{align}
Note that $\langle e_{i} | \delta e_{i} \rangle = 0$ since $|e_{i} \rangle$ and $|\delta e_{i} \rangle$ are belonging to mutually orthogonal state spaces.
 
\subsection{Computation of $\mathcal{N}_{\mathrm{I|E}}(\rho)$}

Using the form of $\rho$ in Eq~\eqref{densitymatrixform} and $|E_{i} \rangle$ in Eq.~\eqref{Decomposition1}, we write the density matrix $\rho$ as
\begin{align}\label{density_matrix}
\rho = \sum_{i=1}^{N} w_{i} |E_{i} \rangle \langle E_{i}| = \sum_{i=1}^{N} w_{i} \Big(|e_{i} \rangle \langle e_{i} | + |e_{i} \rangle \langle \delta e_{i} | + |\delta e_{i} \rangle \langle e_{i} | + |\delta e_{i} \rangle \langle \delta e_{i} |\Big).
\end{align}
To derive $\mathcal{N}_{\mathrm{I|E}}(\rho)$ in Eq.~(6) of the main text, we need to compute $\Vert \rho^{\mathrm{T}_\textrm{I}} \Vert$, which is the sum of the absolute values of all eigenvalues of $\rho^{\mathrm{T}_\textrm{I}}$. For the purpose, we decompose $\rho^{\mathrm{T}_\textrm{I}} = D + F$ into a block diagonal part $D$ and a block off-diagonal part $F$ with the blocks composed of the basis $\{|{\uparrow} \rangle |\phi_{i \uparrow} \rangle,|{\uparrow} \rangle |\phi_{i \downarrow} \rangle,|{\downarrow} \rangle |\phi_{i \uparrow} \rangle,|{\downarrow} \rangle |\phi_{i \downarrow} \rangle\}$.
We notice that $\sum_{i=1}^{N} w_{i} (|e_{i} \rangle \langle e_{i}|)^{\mathrm{T}_\textrm{I}}$ is a block diagonal matrix contributing to $D$, while  $\sum_{i=1}^{N} w_{i} (|e_{i} \rangle \langle \delta e_{i} | + |\delta e_{i} \rangle \langle e_{i} |)^{\mathrm{T}_\textrm{I}}$ is a block off-diagonal matrix contributing to $F$. $(|\delta e_{i} \rangle \langle \delta e_{i} |)^{\mathrm{T}_\textrm{I}}$ contributes to both of $D$ and $F$. Since $\Vert |\delta e_{i} \rangle \Vert = O(E^{\Delta_{z}}) + O(E^{\Delta_{-}})$, the term $(|\delta e_{i} \rangle \langle \delta e_{i} |)^{\mathrm{T}_\textrm{I}}$ is of order $O(E^{2\Delta_{z}}) + O(E^{2\Delta_{-}}) + O(E^{\Delta_{z} + \Delta_{-}}) = O(E^{\mathrm{min}\{2\Delta_{x},2\Delta_{y},2\Delta_{z}\}})$. Therefore, we find
\begin{align}
D & = \sum_{i=1}^{N} w_{i} (|e_{i} \rangle \langle e_{i}|)^{\mathrm{T}_\textrm{I}} + O(E^{\mathrm{min}\{2\Delta_{x},2\Delta_{y},2\Delta_{z}\}}), \label{D_explicit} \\
F & = \sum_{i=1}^{N} w_{i} (|e_{i} \rangle \langle \delta e_{i} | + |\delta e_{i} \rangle \langle e_{i} |)^{\mathrm{T}_\textrm{I}} + O(E^{\mathrm{min}\{2\Delta_{x},2\Delta_{y},2\Delta_{z}\}}). \label{F_explicit}
\end{align}
Since $\Vert D \Vert = O(1)$ and $\Vert F \Vert = O(E^{\Delta_{z}}) + O(E^{\Delta_{-}})$, we compute the eigenvalues $\mu_{i}^{(j)}$'s of $\rho^{\mathrm{T}_\textrm{I}}$ by treating $F$ as a perturbation to $D$.
The eigenvalues of the block diagonal matrix $D$ is determined in each block $i$ and we let $\{\lambda_{i}^{(1)},\lambda_{i}^{(2)},\lambda_{i}^{(3)},\lambda_{i}^{(4)}\}$ be the eigenvalues of the block $i$ of $D$ composed of $\{|{\uparrow} \rangle |\phi_{i \uparrow} \rangle,|{\uparrow} \rangle |\phi_{i \downarrow} \rangle,|{\downarrow} \rangle |\phi_{i \uparrow} \rangle,|{\downarrow} \rangle |\phi_{i \downarrow} \rangle\}$. We below find $\lambda_{i}^{(j)}$ and then obtain $\mu_{i}^{(j)}$. 
Then the trace norm of $\rho^{\mathrm{T}_\textrm{I}}$ is written as $\Vert \rho^{\mathrm{T}_\textrm{I}} \Vert = \sum_{i=1}^{N} \sum_{j=1}^{4} |\mu_{i}^{(j)}|$.

We first find $\lambda_{i}^{(j)}$. For the purpose, we consider the state $|e_{i}' \rangle$ satisfying
$|e_{i} \rangle = |e_{i}' \rangle + O(E^{\mathrm{min}\{2\Delta_{x},2\Delta_{y},2\Delta_{z}\}})$,
\begin{equation}\label{Decomposition4}
\begin{aligned}
|e_{i}' \rangle &= \sqrt{\frac{1 + 2 \langle E_{i} | S_{\mathrm{imp}}^{z} | E_{i} \rangle}{2}} |{\uparrow} \rangle |\phi_{i \uparrow} \rangle \\
&\quad \quad + \sqrt{\frac{2}{1 + 2 \langle E_{i} | S_{\mathrm{imp}}^{z} | E_{i} \rangle}} \langle E_{i} | S_{\mathrm{imp}}^{-} | E_{i} \rangle^{*}  |{\downarrow} \rangle |\phi_{i \uparrow} \rangle + \sqrt{\frac{1 - 4 \langle E_{i} | S_{\mathrm{imp}}^{z} | E_{i} \rangle^{2} - 4 |\langle E_{i} | S_{\mathrm{imp}}^{-} | E_{i} \rangle|^{2}}{2(1 + 2 \langle E_{i} | S_{\mathrm{imp}}^{z} | E_{i} \rangle)}} |{\downarrow} \rangle |\phi_{i \downarrow} \rangle.
\end{aligned}
\end{equation}
See Eq.~\eqref{Decomposition2}. It satisfies $\langle e_{i}' | S_{\mathrm{imp}}^{z} |e_{i}' \rangle = \langle E_{i} | S_{\mathrm{imp}}^{z} | E_{i} \rangle$ and $\langle e_{i}' | S_{\mathrm{imp}}^{-} |e_{i}' \rangle = \langle E_{i} | S_{\mathrm{imp}}^{-} | E_{i} \rangle$. 
We let $\{\sigma_{i}^{(1)},\sigma_{i}^{(2)},\sigma_{i}^{(3)},\sigma_{i}^{(4)}\}$ be eigenvalues of $w_{i} (|e_{i}' \rangle \langle e_{i}'|)^{\mathrm{T}_\textrm{I}}$, and we use Eq.~\eqref{pure_negativity} to have
 \begin{align}\label{Baretracenorm}
\sum_{j=1}^{4} |\sigma_{i}^{(j)}| = w_{i} \Big(\mathcal{N}_{\mathrm{I|E}}(|e_{i}' \rangle) + 1\Big) = w_{i} \Big(\sqrt{1 - 4 \langle e_{i}' | \vec{S}_{\mathrm{imp}} | e_{i}' \rangle^{2}} + 1\Big) = w_{i} \Big(\sqrt{1 - 4 \langle E_{i} | \vec{S}_{\mathrm{imp}} | E_{i} \rangle^{2}} + 1\Big) = w_{i} \Big(\mathcal{N}_{\mathrm{I|E}}(|E_i \rangle) + 1\Big).
\end{align}
And the relation $|e_{i} \rangle = |e_{i}' \rangle + O(E^{\mathrm{min}\{2\Delta_{x},2\Delta_{y},2\Delta_{z}\}})$ and Eq.~\eqref{D_explicit} imply that $\lambda_{i}^{(j)}$ deviates from $\sigma_{i}^{(j)}$ by $O(E^{\mathrm{min}\{2\Delta_{x},2\Delta_{y},2\Delta_{z}\}})$,
\begin{align}\label{Perturbation1}
\lambda_{i}^{(j)} = \sigma_{i}^{(j)} + O(E^{\mathrm{min}\{2\Delta_{x},2\Delta_{y},2\Delta_{z}\}}).
\end{align}

Next we apply the perturbation of $F$ to $\{\lambda_{i}^{(j)}\}$. Since $F$ is a block off-diagonal matrix with respect to the basis $\cup_{i=1}^{N} \{|{\uparrow} \rangle |\phi_{i \uparrow} \rangle,|{\uparrow} \rangle |\phi_{i \downarrow} \rangle,|{\downarrow} \rangle |\phi_{i \uparrow} \rangle,|{\downarrow} \rangle |\phi_{i \downarrow} \rangle\}$, we have
\begin{align}
\Bigg(\sum_{\mu,\eta = \uparrow,\downarrow} c_{\mu \eta}^{*} \langle \mu | \langle \phi_{i \eta}|\Bigg) F \Bigg(\sum_{\mu,\eta = \uparrow,\downarrow} c_{\mu \eta} |\mu \rangle |\phi_{i\eta} \rangle\Bigg) = 0, \quad \quad \forall i, \quad \forall (c_{\mu\eta})_{\mu,\eta = \uparrow,\downarrow} \in \mathbb{C}.
\end{align}
Therefore, the first order contribution of the perturbation $F$ vanishes.
Since $\Vert F \Vert = O(E^{\Delta_{z}}) + O(E^{\Delta_{-}})$, the second order contribution is of order $O(E^{2\Delta_{z}}) + O(E^{2\Delta_{-}}) + O(E^{\Delta_{z} + \Delta_{-}})$.  {In this second order contribution, all nonzero $O(E^{\Delta_{z} + \Delta_{-}})$ terms appear together with nonzero $O(E^{2\Delta_{z}})$ and $O(E^{2\Delta_{-}})$ terms.
Therefore, each eigenvalue $\mu_{i}^{(j)}$ of $\rho^{\mathrm{T}_\textrm{I}}$ is given by
\begin{align}\label{Perturbation2}
\mu_{i}^{(j)} = \lambda_{i}^{(j)} + O(E^{\mathrm{min}\{2\Delta_{x},2\Delta_{y},2\Delta_{z}\}}). 
\end{align}
   
Using Eq.~\eqref{Baretracenorm}, Eq.~\eqref{Perturbation1} and Eq.~\eqref{Perturbation2}, the trace norm $\Vert \rho^{\mathrm{T}_\textrm{I}} \Vert$ becomes
\begin{align}\label{mixedtracenorm}
\Vert \rho^{\mathrm{T}_\textrm{I}} \Vert 
&= \sum_{i=1}^{N} \sum_{j=1}^{4} |\sigma_{i}^{(j)}| + O(E^{\mathrm{min}\{2\Delta_{x},2\Delta_{y},2\Delta_{z}\}}) \nonumber \\
&= \sum_{i=1}^{N} w_{i} \Big(\mathcal{N}_{\mathrm{I|E}}(|E_{i} \rangle) + 1 \Big) + O(E^{\mathrm{min}\{2\Delta_{x},2\Delta_{y},2\Delta_{z}\}}).
\end{align}
The identification $E_i \sim T$, Eq.~\eqref{mixedtracenorm} with $w_{i} = e^{-E_{i}/T}/Z$, and $\sum_{i=1}^{N} w_{i} = 1$ lead to
\begin{align}
\mathcal{N}_{\mathrm{I|E}}(\rho) = \Vert \rho^{\mathrm{T}_\textrm{I}} \Vert - 1 
&= \sum_{i=1}^{N} w_{i}(E_{i}) \mathcal{N}_{\mathrm{I|E}}(|E_{i} \rangle)|_{E_{i} \sim T} + O(T^{\mathrm{min}\{2\Delta_{x},2\Delta_{y},2\Delta_{z}\}}),
\end{align}
which is Eq.~(6) of the main text.


\section{VI. NRG calculation and parameters}

We compute the entanglement negativity $\mathcal{N}_\textrm{I$|$E}$ between the impurity and channels in the multichannel Kondo model by using the NRG. The detailed way of computing the negativity based on the NRG has been developed in Ref.~\cite{Shim18_supp}.

We describe the parameters used in the NRG calculation.
Each channel has a constant density of states $1/(2D)$ within $[-D,D]$.
We choose the parameters as follows: the half band width $D$ is 1,
the Kondo coupling is 0.3$D$,
the discretization parameter $\Lambda$ is 10,
and the length of each Wilson chain is 28.
The number of kept states is chosen as 300 for the 1CK model, 3,000 for the 2CK model, and 10,000 for the 3CK model.
To solve the multichannel Kondo model, we employ the interleaved NRG~\cite{Mitchell14_supp,Stadler16_supp} for spin and channel indices.
The $z$-averaging is done with the two values of $z=0$ and $1/2$ when the entanglement negativity is computed,
and with the single value of $z=0$ when the matrix elements of the impurity spin is computed.

The NRG calculation of $\mathcal{N}_\textrm{I$|$E}$ agrees with Eq.~(1) as shown in Fig.~(2c).
Here we find $a_{k}\sim 0.7$, $0.4$, $0.3$ for $k=1,2,3$, respectively, and $T_\mathrm{K} \sim 4.93\times 10^{-4}D$ by using the Poor man's scaling.
  
\section{VII. NRG calculation of matrix elements of the impurity spin operator}
We compute matrix elements $|\langle E|S_\text{imp}^{\alpha = z,+}|E'\rangle|$ of the impurity spin $\vec{S}_\text{imp}$ for energy eigenstates $|E\rangle$ and $|E'\rangle$ by using the NRG.
At the $n$th NRG iteration step,
we compute $|\langle E_{ni}^K |S_\text{imp}^{\alpha=z,+}|E_{ni'}^D \rangle|$, $|\langle E_{ni}^D |S_\text{imp}^{\alpha=z,+}|E_{ni'}^K \rangle|$,  and $|\langle E_{ni}^D |S_\text{imp}^{\alpha=z,+}|E_{ni'}^D \rangle|$, where $|E_{ni}^{K(D)} \rangle$ denotes the $i$th kept (discarded) energy eigenstate at the $n$th NRG iteration step.
At each NRG iteration step we select the 100 number of the largest ones among the computed matrix elements; the number is turned out to be sufficiently large, as we achieved convergence of the results with increasing the number.
We collect the selected matrix elements for all the NRG iteration steps, and plot them in Fig.~\ref{MCKE_supp_fig2} with respect to the average energy $(E+E')/2$ in the unit of $T_\text{K}$.

Figure~\ref{MCKE_supp_fig2} shows that
the matrix elements obey a power-law scaling behavior in the average energy $(E+E')/2$ at low energy $E+E' \ll T_\text{K}$.
The power law exponent is 1 for the 1CK model,
1/2 for the 2CK,
and 2/5 for the 3CK.
These agree with the BCFT result  $\langle E_i|S_\text{imp}^{\alpha = x,y,z}|E_j\rangle \propto E^\Delta$ with $E \sim E_i \sim E_j \ll T_\text{K}$ in Eq.~(10) of the main text.
 
\begin{figure}
\centerline{\includegraphics[width=1\textwidth]{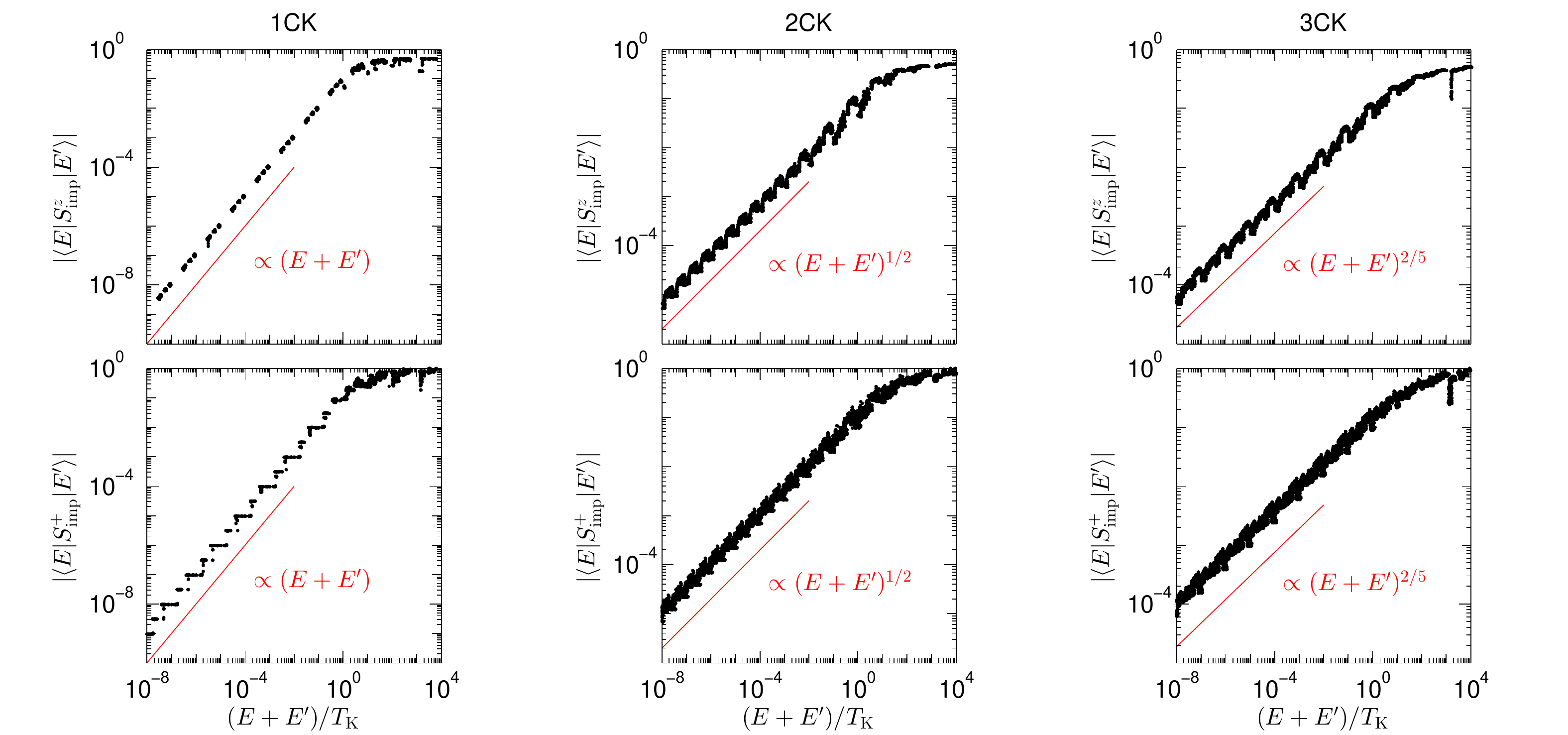}}
\caption{Matrix elements $|\langle E|S_\text{imp}^{\alpha = z,+}|E'\rangle|$ of the impurity spin $\vec{S}_\text{imp}$ with respect to the energy eigenstates $|E\rangle$ and $|E'\rangle$ computed by using the NRG.
$|\langle E|S_\text{imp}^{\alpha = z,+}|E'\rangle|$'s obey a power law scaling in the average energy $(E+E')/2$ for low energies $E$, $E' \ll T_\text{K}$.
The scaling exponent is 1 for the 1CK model, 1/2 for the 2CK, and 2/5 for the 3CK.
}\label{MCKE_supp_fig2}
\end{figure}

\section{VIII. Derivation of Eq.~(11)}

We derive Eq.~(11) in the main text.
We consider a restricted Hilbert space spanned by states of energy $\sim E$. We start from a general form of $|E_{i} \rangle$, $|E_{i} \rangle = |{\uparrow} \rangle  |\widetilde{\psi}_{i{\uparrow}} \rangle + |{\downarrow} \rangle |\widetilde{\psi}_{i{\downarrow}} \rangle$, where $|\mu = {\uparrow},{\downarrow} \rangle$'s are the impurity spin eigenstates of $S_{\mathrm{imp}}^{z}$, and $|\widetilde{\psi}_{i{\mu}} \rangle$'s are unnormalized channel states.
The normalization $\langle E_{i} | E_{j} \rangle = \delta_{ij}$ and the scaling behavior $\langle E_{i} | S_{\mathrm{imp}}^{\alpha} |E_{j} \rangle = O((E / T_{K})^{\Delta})$ of the impurity spin yield
\begin{equation}\label{scaling_supp}
\begin{aligned}
\langle \widetilde{\psi}_{i {\uparrow}} | \widetilde{\psi}_{j {\uparrow}} \rangle + \langle \widetilde{\psi}_{i{\downarrow}} | \widetilde{\psi}_{j{\downarrow}} \rangle &= \delta_{ij}, &&\qquad \forall i,j \\
\langle \widetilde{\psi}_{i {\uparrow}} | \widetilde{\psi}_{j {\uparrow}} \rangle - \langle \widetilde{\psi}_{i{\downarrow}} | \widetilde{\psi}_{j{\downarrow}} \rangle &= O((E / T_{K})^{\Delta}), &&\qquad \forall i,j \\
\langle \widetilde{\psi}_{i {\uparrow}} | \widetilde{\psi}_{j {\downarrow}} \rangle &= O((E / T_{K})^{\Delta}) &&\qquad \forall i,j.
\end{aligned}
\end{equation}
By applying the Gram-Schmidt process to $|\widetilde{\psi}_{i \mu} \rangle$'s, we introduce the states $| \phi_{j\eta} \rangle$'s satisfying $\langle \phi_{i\mu} | \phi_{j\eta} \rangle = \delta_{ij} \delta_{\mu \eta}$  as in Eqs.~\eqref{Gram_Schmidt5-1}-\eqref{Gram_Schmidt5-4}.
Then, each $|\widetilde{\psi}_{i \mu} \rangle$ is written as $|\widetilde{\psi}_{i\mu} \rangle = [(1 + a_{i\mu}) |\phi_{i\mu} \rangle + |\chi_{i\mu} \rangle]/\sqrt{2}$ in terms of a coefficient $a_{i\mu}$ (related with normalization) and a unnormalized state $|\chi_{i\mu} \rangle$ orthogonal to $|\phi_{i\mu}\rangle$ (namely, $\langle \phi_{i\mu} | \chi_{i\mu} \rangle = 0$), and $|E_{i} \rangle = |{\uparrow} \rangle  |\widetilde{\psi}_{i{\uparrow}} \rangle + |{\downarrow} \rangle  |\widetilde{\psi}_{i{\downarrow}} \rangle$ are consequently expressed as 
\begin{align}
|E_{i} \rangle &= \frac{1}{\sqrt{2}} \Bigg(|{\uparrow} \rangle \otimes \Big((1 + a_{i {\uparrow}}) |\phi_{i {\uparrow}} \rangle + |\chi_{i {\uparrow}} \rangle\Big) + |{\downarrow} \rangle \otimes \Big((1 + a_{i {\downarrow}}) |\phi_{i{\downarrow}} \rangle + |\chi_{i{\downarrow}} \rangle\Big)\Bigg) \nonumber
\\
&= \underbrace{\frac{1}{\sqrt{2}} \Big[|{\uparrow} \rangle \otimes |\phi_{i {\uparrow}} \rangle + |{\downarrow} \rangle \otimes |\phi_{i{\downarrow}} \rangle\Big]}_{:\text{maxiamlly entangled} \quad (\because \{|\phi_{i\eta} \rangle\} \text{ are orthonormal})} + \underbrace{\frac{1}{\sqrt{2}} \Bigg[|{\uparrow} \rangle \otimes \Big(a_{i {\uparrow}} |\phi_{i {\uparrow}} \rangle + |\chi_{i {\uparrow}} \rangle\Big) + |{\downarrow} \rangle \otimes \Big(a_{i{\downarrow}} |\phi_{i{\downarrow}} \rangle + |\chi_{i{\downarrow}} \rangle\Big)\Bigg]}_{= |\delta E_{i} \rangle}, \label{expliciteig}
\end{align}
which is the form in Eq.~(11) of the main text.

We combine Eq. (\ref{scaling_supp}) with the relations of $a_{j\eta}$ and $|\chi_{j\eta} \rangle$
\begin{equation}
a_{j\eta} = \sqrt{2} \langle \phi_{j\eta} | \widetilde{\psi}_{j\eta} \rangle - 1, \qquad |\chi_{j\eta} \rangle = \sqrt{2} (|\widetilde{\psi}_{j\eta} \rangle - |\phi_{j\eta} \rangle \langle \phi_{j\eta} | \widetilde{\psi}_{j\eta} \rangle)
\end{equation}
to obtain the following estimates 
\begin{equation}\label{scaling2_supp}
\begin{aligned}
a_{j\eta} &= O((E / T_{K})^{\Delta})  &&\qquad \forall (j,\eta), \\
\langle \phi_{i\mu} | \chi_{j\eta} \rangle &= O((E / T_{K})^{\Delta}) &&\qquad \forall (i,\mu),(j,\eta) \text{ with }(i,\mu) \neq (j,\eta), \\
\langle \chi_{i \mu} | \chi_{j \eta} \rangle &= O((E / T_{K})^{2 \Delta})  &&\qquad \forall (i,\mu),(j,\eta).
\end{aligned}
\end{equation}
We note that $a_{j \eta}$'s are taken as real numbers in the Gram-Schmidt process.
Using the estimates, we find that $|\delta E_i\rangle$ has the norm of $O((E/T_\text{K})^\Delta)$:
\begin{align}
\sqrt{\langle\delta E_i|\delta E_i\rangle} &= \sqrt{\frac{1}{2} \Big[a_{i\uparrow}^{2} + \langle\chi_{i\uparrow}|\chi_{i\uparrow}\rangle + a_{i\downarrow}^{2} + \langle\chi_{i\downarrow}|\chi_{i\downarrow}\rangle\Big]} \nonumber
\\
&= \sqrt{O((E/T_\text{K})^{2\Delta})} = O((E/T_\text{K})^\Delta). \\
\langle\delta E_i| E_j\rangle &= \frac{1}{2} \sum_{\mu = \uparrow,\downarrow} \Big[a_{i\mu}(1 + a_{j \mu}) \delta_{ij} + a_{i\mu} \langle \phi_{i\mu} | \chi_{j\mu} \rangle + (1 + a_{j \mu}) \langle \chi_{i \mu} | \phi_{j \mu} \rangle + \langle \chi_{i\mu} | \chi_{j \mu} \rangle\Big] \nonumber
\\
&= O((E/T_\text{K})^\Delta).
\end{align}

\section{IX. Direct calculation of $\Vert \rho^{\mathrm{T}_\textrm{I}} \Vert = 2 - O((\frac{T}{T_{\mathrm{K}}})^{2 \Delta})$}

We derive $\Vert \rho^{\mathrm{T}_\textrm{I}} \Vert = 2 - O((\frac{T}{T_{\mathrm{K}}})^{2 \Delta})$ of the main text.
Using the form of $\rho$ in Eq~\eqref{densitymatrixform} and $|E_{i} \rangle$ in Eq.~\eqref{expliciteig}, we write the density matrix $\rho$ 
\begin{align}
\rho & = \sum_{j,j'} \sum_{\mu,\mu',\eta,\eta' = {\uparrow},{\downarrow}} \rho_{(\mu,j,\eta),(\mu',j',\eta')} |\mu \rangle \langle \mu'| \otimes |\phi_{j \eta} \rangle \langle \phi_{j'\eta'}| , \\
\rho_{(\mu,j,\eta),(\mu',j',\eta')}  & = \sum_{i} \frac{w_{i}}{2} \Big[(1 + a_{i\mu}) \delta_{ij} \delta_{\mu\eta} + \langle \phi_{j\eta} | \chi_{i\mu} \rangle\Big]\Big[(1 + a_{i\mu'}) \delta_{ij'} \delta_{\mu'\eta'} + \langle \chi_{i\mu'} | \phi_{j'\eta'} \rangle\Big].
\label{elementrho_supp}
\end{align}
Applying the partial transpose to $\rho$ in Eq.~(\ref{elementrho_supp}) yields
\begin{align}
\rho^{\mathrm{T}_\textrm{I}} = \sum_{j,j'} \sum_{\mu,\mu',\eta,\eta' = {\uparrow},{\downarrow}} \rho_{(\mu',j,\eta),(\mu,j',\eta')} |\mu \rangle \langle \mu'| \otimes |\phi_{j \eta} \rangle \langle \phi_{j'\eta'}| .
\end{align}
Since $\Vert \rho^{\mathrm{T}_\textrm{I}} \Vert$ is the sum of the absolute values of all eigenvalues of $\rho^{\mathrm{T}_\textrm{I}}$, $\Vert \rho^{\mathrm{T}_\textrm{I}} \Vert$ is equal to the sum of the square roots of all eigenvalues of $(\rho^{\mathrm{T}_\textrm{I}})^{2}$ which becomes
\begin{align}
(\rho^{\mathrm{T}_\textrm{I}})^{2} = \sum_{j,j'} \sum_{\mu,\mu',\eta,\eta' = \{{\uparrow},{\downarrow}\}} \Bigg(\sum_{n} \sum_{\nu,\zeta = {\uparrow},{\downarrow}} \rho_{(\nu,j,\eta),(\mu,n,\zeta)} \rho_{(\mu',n,\zeta),(\nu,j',\eta')}\Bigg) |\mu \rangle \langle \mu' | \otimes |\phi_{j\eta} \rangle \langle \phi_{j'\eta'}| .
\end{align}
We decompose $(\rho^{\mathrm{T}_\textrm{I}})^{2}$ into the diagonal part $D$ and the off-diagonal part $F$, $(\rho^{\mathrm{T}_\textrm{I}})^{2} = D + F$, where their matrix elements are
\begin{align}\label{elementD_supp}
D_{(\mu,j,\eta),(\mu',j',\eta')} &= \frac{w_{j}^{2}}{4} \Big[1 + 2 a_{j\eta} + 2 a_{j \mu} + O((T / T_{K})^{2 \Delta})\Big] \delta_{\mu\mu'} \delta_{jj'} \delta_{\eta\eta'} ,
\end{align}
and
\begin{align}\label{elementF_supp}
F_{(\mu,j,\eta),(\mu',j',\eta')} &= \frac{w_{j'}^{2}}{4} \langle \phi_{j\eta} | \chi_{j'\eta'} \rangle \delta_{\mu\mu'} + \frac{w_{j} w_{j'}}{4} \langle \chi_{j\mu} | \phi_{j'\mu'} \rangle \delta_{\eta\eta'} \nonumber \\
&\qquad \qquad + \frac{w_{j} w_{j'}}{4} \langle \phi_{j\mu} | \chi_{j'\mu'} \rangle \delta_{\eta\eta'} + \frac{w_{j}^{2}}{4} \langle \chi_{j\eta} | \phi_{j'\eta'} \rangle \delta_{\mu\mu'} + O((T / T_{K})^{2 \Delta}) \nonumber \\
&= \big[1-\delta_{\mu\mu'}\delta_{jj'}\delta_{\eta\eta'}\big] O((T / T_\text{K})^{\Delta}) .
\end{align}
Since  $F \propto O((T / T_{K})^{\Delta})$,
we can treat $F$ as a perturbation in computing the eigenvalues of $(\rho^{\mathrm{T}_\textrm{I}})^{2}$.
When there is no degeneracy in the values of $D_{(\mu,j,\eta),(\mu,j,\eta)}$'s, we apply the non-degenerate perturbation theory to compute the eigenvalues $\overline{\sigma}_{\mu j \eta}$ of $(\rho^{\mathrm{T}_\textrm{I}})^{2}$:
\begin{align}
\overline{\sigma}_{\mu j \eta} &= D_{(\mu,j,\eta),(\mu,j,\eta)} + (\langle \mu | \otimes \langle \phi_{j\eta}|) F (| \mu \rangle \otimes |\phi_{j\eta} \rangle) + \sum_{|\nu \rangle \otimes |\phi_{n\zeta} \rangle \neq |\mu \rangle \otimes |\phi_{j\eta} \rangle} \frac{|(\langle \nu | \otimes \langle \phi_{n\zeta}|) F (|\mu \rangle \otimes |\phi_{j\eta} \rangle)|^{2}}{D_{(\mu,j,\eta),(\mu,j,\eta)} - D_{(\nu,n,\zeta),(\nu,n,\zeta)}} + O((T/T_{K})^{3\Delta}) \nonumber \\
&= D_{(\mu,j,\eta),(\mu,j,\eta)} + f_{\mu j \eta}(T / T_\text{K})^\Delta + O((T / T_\text{K})^{2 \Delta})  \nonumber \\
&= D_{(\mu,j,\eta),(\mu,j,\eta)} + O((T / T_\text{K})^{2 \Delta}). \label{nondegeneratePT}
\end{align}
for each index $(\mu,j,\eta)$. In the second equality, the term of the order of $(T / T_\text{K})^\Delta$ comes from
the first order perturbation on $F$, $f_{\mu j \eta}(T / T_\text{K})^\Delta =  (\langle \mu | \otimes \langle \phi_{j\eta}|) F (| \mu \rangle \otimes |\phi_{j\eta} \rangle)$.
This term vanishes, $f_{\mu j \eta} = 0$, for each index $(\mu, j, \eta)$, since $F$ is an off-diagonal matrix.
On the other hand, when there is degeneracy in the values of $D_{(\mu,j,\eta),(\mu,j,\eta)}$'s, namely when
$D_{(\mu,j,\eta),(\mu,j,\eta)} = D$ for some $(\mu,j,\eta)$'s,
we apply the degenerate perturbation theory, to compute the eigenvalues $\overline{\sigma}_{a}$ of $(\rho^{\mathrm{T}_\textrm{I}})^{2}$ in the degeneracy manifold of the $(\mu,j,\eta)$'s:
\begin{align}
\overline{\sigma}_{a} &= D + \langle \Psi_{a}^{(n)} | F | \Psi_{a}^{(n)} \rangle+ O((T / T_\text{K})^{2 \Delta})  \nonumber \\
&= D  + f_{a}(T / T_\text{K})^\Delta + O((T / T_\text{K})^{2 \Delta}). \label{degeneratePT}
\end{align}
Here, $a$ is the index for counting the states $|\mu \rangle \otimes |\phi_{j \eta} \rangle$ in the degeneracy manifold, and $\{|\Psi_{a}^{(n)} \rangle\}_{a}$ is the set of the states digonalizing the matrix $F_M$ which is the matrix $F$ projected onto the manifold. We find an interesting relation used later,
\begin{align}
\sum_{a} f_{a}(T / T_{K})^{\Delta} = \sum_{a} \langle \Psi_{a}^{(n)} | F | \Psi_{a}^{(n)} \rangle = \mathrm{tr}(F_M) = 0 \label{degeneratePT2}
\end{align}
for each degeneracy manifold. The last equality comes from the fact that $F$ (hence $F_M$) is off-diagonal so that its trace is zero.

 
We calculate $\Vert \rho^{\mathrm{T}_\textrm{I}} \Vert$ as the sum of the square roots of the eigenvalues $\bar{\sigma}_{\mu j\eta}$ of $(\rho^{\mathrm{T}_\textrm{I}})^{2}$.
We represent the eigenvalue $\bar{\sigma}_{\mu j\eta}$ as
\begin{align}
\bar{\sigma}_{\mu j \eta} = \frac{w_{j}^{2}}{4}(1 + 2 a_{j\eta} + 2 a_{j\mu}) + f_{\mu j \eta}(T / T_{K})^{\Delta} + O((T / T_{K})^{2 \Delta}) .
\end{align}
Then we do the Taylor expansion such that
\begin{align}
\Vert \rho^{\mathrm{T}_\textrm{I}} \Vert = \sum_{j}\sum_{\eta,\mu={\uparrow},{\downarrow}} \sqrt{\bar{\sigma}_{\mu j\eta}}
&= \sum_{j} \sum_{\eta,\mu = {\uparrow},{\downarrow}} \sqrt{\frac{w_{j}^{2}}{4} (1 + 2 a_{j \eta} + 2 a_{j\mu}) + f_{\mu j \eta} (T / T_{K})^{\Delta} + O((T / T_{K})^{2 \Delta})} \nonumber \\
&= \sum_{j} \sum_{\eta,\mu = {\uparrow},{\downarrow}} \frac{w_{j}}{2} (1 + a_{j \eta} + a_{j\mu}) + \frac{1}{w_{j}} f_{\mu j \eta}(T / T_{K})^{\Delta} + O((T / T_{K})^{2 \Delta}) \nonumber \\
&= \sum_{j} \sum_{\eta,\mu = {\uparrow},{\downarrow}} \frac{w_{j}}{2} (1 + a_{j \eta} + a_{j\mu}) + O((T / T_{K})^{2 \Delta}) . \label{tracenorm_intermediate}
\end{align}
In the last equality, we use the facts: (1) $f_{\mu j \eta} = 0$  in Eq.~\eqref{nondegeneratePT} for the non-degenerate case,
and (2)  $\sum_a f_a = 0$ in Eq.~\eqref{degeneratePT2} and $w_j$ is constant in each degenerate manifold.
Because $a_{j{\uparrow}} + a_{j{\downarrow}} = 0$, which can be proved by applying the normalization $\langle E_{i} | E_{i} \rangle = 1$ to Eq.~\eqref{expliciteig}, we find that Eq.~\eqref{tracenorm_intermediate} becomes
\begin{align}
\Vert \rho^{\mathrm{T}_\textrm{I}} \Vert &= \sum_{j} \sum_{\eta,\mu = {\uparrow},{\downarrow}} \frac{w_{j}}{2} + O((T / T_{K})^{2 \Delta}) = 2 - O((T / T_{K})^{2 \Delta}),
\end{align}
which is $\Vert \rho^{\mathrm{T}_\textrm{I}} \Vert = 2 - O((\frac{T}{T_{\mathrm{K}}})^{2 \Delta})$ of the main text. In the last equality, we have used the probability conservation of $\sum_j w_j = 1$ and $\Vert \rho^{\mathrm{T}_\textrm{I}} \Vert \leq 2$.


\section{IX. Derivation of Eq.~(13)}

To derive Eq.~(14) in the main text, 
we solve exactly the anisotropic 2CK model by using the bosonization.
We consider the following anisotropic 2CK model Hamiltonian
\begin{align}
H &= \sum_{i=1,2} \Big[H_{i} + \lambda_{z} S_{\mathrm{imp}}^{z} S_{i}^{z} + \sum_{\alpha = +,-} \frac{\lambda_{\perp} + (-1)^{i} \delta \lambda_{\perp}}{2} S_{\mathrm{imp}}^{\alpha} S_{i}^{-\alpha}\Big]
\end{align}
where $H_{i}$ is the noninteracting Hamiltonian of the $i$th channel, $\lambda_{z} S_{\mathrm{imp}}^{z} S_{i}^{z} + \frac{\lambda_{\perp}}{2} S_{\mathrm{imp}}^{+} S_{i}^{-} + \frac{\lambda_{\perp}}{2} S_{\mathrm{imp}}^{-} S_{i}^{+}$ is the Kondo interaction, $\delta \lambda_{\perp}$ introduces the anisotropic coupling,
$\vec{S}_{\mathrm{imp}}$ is the impurity spin, $\vec{S}_{i} = \sum_{\alpha,\alpha' = +,-} : \psi_{\alpha i}^{\dagger}(0) \frac{1}{2} \vec{\sigma}_{\alpha \alpha'} \psi_{\alpha' i}(0) :$ is the $i$th channel electron spin at the impurity position with $S_{\text{imp}/i}^{\pm} = S_{\text{imp}/i}^{x} \pm i S_{\text{imp/i}}^{y}$, $\vec{\sigma}$ is a Pauli matrix, $\psi_{\alpha i}(x)$ is the electron field with spin $\alpha = (+,-)$ and channel $i = (1,2)$ at position $x$,
and :: refers the normal ordering.
By using the bosonization and the Emery-Kivelson transformation at the Toulouse point~\cite{Zarand00_supp}, the above Hamiltonian is mapped into $\widetilde{H} = e^{iS_{\mathrm{imp}}^{z} \varphi_{s}(0)} H e^{-iS_{\mathrm{imp}}^{z} \varphi_{s}(0)}$,
where $\varphi_{s}(0)$ is a boson field of the spin degree of freedom.
For the energy eigenstate $|E \rangle$ of the original Hamiltonian $H$,
we consider an eigenstate of $\widetilde{H}$, $|\widetilde{E} \rangle = e^{i S_{\mathrm{imp}}^{z} \varphi_{s}(0)} |E \rangle$.
By the refermionization~\cite{Zarand00_supp}, the transformed Hamiltonian $\widetilde{H}$ is written as $\widetilde{H} = \sum_{y=c,s,f,x}H_y + E_G + \textrm{constant}$ where $c$, $s$, $f$ and $x$ denotes the charge, spin, flavor, and spin-flavor degrees of freedom of the channels and $E_G$ is the ground state energy.
Here,
only the spin-flavor degrees of freedom is coupled to the impurity such that
\begin{align}
H_{x} &= \sum_{k} k : c_{k}^{\dagger} c_{k} : + \sqrt{\frac{2 \pi \Gamma}{L}} \sum_{k} (c_{k}^{\dagger} + c_{k})(c_{d} - c_{d}^{\dagger}) + \sqrt{\frac{2 \pi \delta \Gamma}{L}} \sum_{k} (c_{k} - c_{k}^{\dagger})(c_{d} + c_{d}^{\dagger}) ,
\end{align}
where $c_{d}$ is a local pseudofermion at the impurity position,
$c_{k}$ is a pseudofermion with momentum $k$, $\Gamma = \lambda_{\perp}^{2}/4a$, $\delta \Gamma = (\delta \lambda_{\perp})^{2} / 4a$, and $a$ is a short distance cutoff.
The charge, spin, and flavor degrees of freedom are decoupled from the impurity
and so we drop them henceforth.
The impurity spin can be expressed as $S_{\mathrm{imp}}^{z} = c_{d}^{\dagger} c_{d} - 1/2$ and $S_{\mathrm{imp}}^{-} = \mathcal{F}_{s} c_{d}$,
where $\mathcal{F}_{s}$ is the Klein factor corresponding to the spin degree of freedom.
We define the Majorana fermions 
\begin{align}
\begin{cases} \gamma_{d+} = (c_{d} + c_{d}^{\dagger})/ \sqrt{2} \\ \gamma_{d-} = -i (c_{d}^{\dagger} - c_{d})/\sqrt{2} \end{cases}, \quad \begin{cases} \gamma_{k+} = (c_{k} + c_{-k} + c_{k}^{\dagger} + c_{-k}^{\dagger}) / 2 \\ \gamma_{k-} = -i(c_{k} - c_{-k} - c_{k}^{\dagger} + c_{-k}^{\dagger}) / 2 \end{cases}, \quad \begin{cases} \eta_{k+} = -i (c_{k} + c_{-k} - c_{k}^{\dagger} - c_{-k}^{\dagger}) / 2 \\ \eta_{k-} = (- c_{k} + c_{-k} - c_{k}^{\dagger} + c_{-k}^{\dagger}) / 2 \end{cases}
\end{align}
for $k > 0$. The Hamiltonian $H_{x}$ is diagonalized as
\begin{align}
H_{x} = \sum_{\epsilon \geq 0} \epsilon \Big[i \widetilde{\gamma}_{\epsilon +} \widetilde{\gamma}_{\epsilon -} + \frac{1}{2}\Big] + \delta E_{G}
\end{align}
by using the Bogoliubov transformation
\begin{align}
\begin{cases} \widetilde{\gamma}_{\epsilon +}
=
B_{\epsilon d+} \gamma_{d+} + \sum_{k > 0} B_{\epsilon(k,1)+} \gamma_{k+} + \sum_{k > 0} B_{\epsilon (k,2) +} \eta_{k-}
\\
\widetilde{\gamma}_{\epsilon -}
=
B_{\epsilon d-} \gamma_{d-} + \sum_{k > 0} B_{\epsilon(k,1)-} \gamma_{k-} + \sum_{k > 0} B_{\epsilon (k,2) -} \eta_{k+} \end{cases} .
\end{align}
Here, $\delta E_{G}$ is the ground state energy shift
and $\epsilon$ is the excitation energy given by
\begin{align}
\Bigg(\epsilon + 4 \pi \Gamma \tan \frac{\epsilon L}{2}\Bigg) \Bigg(\epsilon + 4 \pi \delta \Gamma \tan \frac{\epsilon L}{2}\Bigg) = 0 .
\end{align}
Hence, the eigenstate $|\widetilde{E} \rangle$ of $\widetilde{H}$ can be expressed a state made by applying $\widetilde{\gamma}_{\epsilon +}$, $\widetilde{\gamma}_{\epsilon -}$, and the creation and annihilation operators of the charge, spin, and flavor degree of freedom to the ground state.
The coefficients $B_{\epsilon d \pm}$ are
\begin{align}
B_{\epsilon d +} &= \begin{cases} \frac{1}{\sqrt{1 + 2 \pi L (\delta \Gamma)}} &\quad \text{for } \epsilon = 0
\\
0 &\quad \text{for } \epsilon \neq 0 \text{ and } \epsilon + 4 \pi \Gamma \tan \frac{\epsilon L}{2} = 0
\\
\Big(\frac{4 \pi \delta \Gamma / L}{\epsilon^{4}/4 + 2 \pi \delta \Gamma \epsilon^{2} / L + 4 \pi^{2} (\delta \Gamma)^{2} \epsilon^{2}}\Big)^{1/2} \epsilon &\quad \text{for }\epsilon \neq 0 \text{ and }\epsilon + 4 \pi \delta \Gamma \tan \frac{\epsilon L}{2} = 0 \end{cases} ,
\\
B_{\epsilon d -} &= \begin{cases} \frac{1}{\sqrt{1 + 2 \pi L \Gamma}} &\quad \text{for } \epsilon = 0
\\
\Big(\frac{4 \pi \Gamma / L}{\epsilon^{4}/4 + 2 \pi \Gamma \epsilon^{2} / L + 4 \pi^{2} \Gamma^{2} \epsilon^{2}}\Big)^{1/2} \epsilon &\quad \text{for }\epsilon \neq 0 \text{ and } \epsilon + 4 \pi \Gamma \tan \frac{\epsilon L}{2} = 0
\\
0 &\quad \text{for } \epsilon \neq 0 \text{ and } \epsilon + 4 \pi \delta \Gamma \tan \frac{\epsilon L}{2} = 0 \end{cases} .
\end{align}
We focus on the two cases $\delta \Gamma \ll 1 / L \ll \Gamma$ and $1/L \ll \Gamma, \delta \Gamma$.
The first one corresponds to $T^* \ll T \ll T_\text{K}$
and the other one corresponds to $T \ll T^*, T_\text{K}$,
as we will replace $1/L \sim T$ (based on the reasoning discussed in the above) and $\delta \Gamma \sim T^*/(\nu T_\text{K})$ at the last.
For each case,
the coefficients become
\begin{align}\label{Bscaling}
B_{\epsilon d +} = \begin{cases} O(1) &\quad \text{for } \delta \Gamma \ll (1 / L) \ll \Gamma \text{ and }\epsilon = 0
\\
O(\sqrt{L \delta \Gamma}) &\quad \text{for } \delta \Gamma \ll (1 / L) \ll \Gamma\text{ and }\epsilon \neq 0
\\
O(1 / \sqrt{L \delta \Gamma}) &\quad \text{for } (1/L) \ll \Gamma, \, \delta \Gamma \end{cases},
\quad  B_{\epsilon d -} = \begin{cases} O(1 / \sqrt{L \Gamma}) &\quad \text{for } \delta \Gamma \ll (1 / L) \ll \Gamma
\\
O(1 / \sqrt{L \Gamma}) &\quad \text{for } (1/L) \ll \Gamma, \, \delta \Gamma \end{cases}.
\end{align}

We now derive Eq.~(14),
the scaling behavior of $S_{\mathrm{imp}}^{z}$ and $S_{\mathrm{imp}}^{-}$.
The orthogonality of the matrix $(B_{\epsilon n \pm})_{n = d,(k,1),(k,2)}$ gives inverse relations $\gamma_{d+} = \sum_{\epsilon} B_{\epsilon d+} \widetilde{\gamma}_{\epsilon +}$ and $\gamma_{d-} = \sum_{\epsilon} B_{\epsilon d-} \widetilde{\gamma}_{\epsilon -}$.
By using these inversion relations and $S_{\mathrm{imp}}^{z} = c_{d}^{\dagger} c_{d} - 1/2 = - i \gamma_{d+} \gamma_{d-}$,
we obtain
\begin{align}\label{Szscaling1}
\langle E_{i} | S_{\mathrm{imp}}^{z} | E_{j} \rangle = \langle \widetilde{E}_{i} | S_{\mathrm{imp}}^{z} | \widetilde{E}_{j} \rangle = \langle \widetilde{E}_{i} | (c_{d}^{\dagger} c_{d} - 1/2) | \widetilde{E}_{j} \rangle = -i \sum_{\epsilon,\epsilon' \geq 0} B_{\epsilon d+} B_{\epsilon' d-} \langle \widetilde{E}_{i} | \widetilde{\gamma}_{\epsilon +} \widetilde{\gamma}_{\epsilon' -} | \widetilde{E}_{j} \rangle.
\end{align}
Applying Eq.~(\ref{Bscaling}) and $\langle \widetilde{E}_{i} | \widetilde{\gamma}_{\epsilon +} \widetilde{\gamma}_{\epsilon' -} | \widetilde{E}_{j} \rangle = O(1)$ to Eq.~(\ref{Szscaling1}) yields
\begin{align}\label{Szscaling2}
\langle E_{i} | S_{\mathrm{imp}}^{z} | E_{j} \rangle = \begin{cases} O(1 / \sqrt{L \Gamma}) &\quad \text{for }\delta \Gamma \ll (1 / L) \ll \Gamma \\  O(1 / L \sqrt{\Gamma \delta \Gamma}) &\quad \text{for }(1/L) \ll \Gamma \, \& \, \delta \Gamma \end{cases}.
\end{align}
Substituting $S_{\mathrm{imp}}^{-} = \mathcal{F}_{s} c_{d}$ gives
\begin{align}\label{S-scaling1}
\langle E_{i} | S_{\mathrm{imp}}^{-} | E_{j} \rangle &= \langle \widetilde{E}_{i} | e^{i S_{\mathrm{imp}}^{z} \varphi_{s}(0)} S_{\mathrm{imp}}^{-} e^{- i S_{\mathrm{imp}}^{z} \varphi_{s}(0)} | \widetilde{E}_{j} \rangle \nonumber \\ 
&= \langle \widetilde{E}_{i} | e^{-i \varphi_{s}(0)} S_{\mathrm{imp}}^{-} | \widetilde{E}_{j} \rangle \nonumber \\
&= \langle \widetilde{E}_{i} | e^{-i \varphi_{s}(0)} \mathcal{F}_{s} c_{d} | \widetilde{E}_{j} \rangle \nonumber \\
&= \frac{1}{\sqrt{2}} \sum_{\epsilon \geq 0} B_{\epsilon d +} \langle \widetilde{E}_{i} | e^{-i\varphi_{s}(0)} \mathcal{F}_{s} \widetilde{\gamma}_{\epsilon +} | \widetilde{E}_{j} \rangle - \frac{i}{\sqrt{2}} \sum_{\epsilon \geq 0} B_{\epsilon d -} \langle \widetilde{E}_{i} | e^{-i\varphi_{s}(0)} \mathcal{F}_{s} \widetilde{\gamma}_{\epsilon -} | \widetilde{E}_{j} \rangle.
\end{align}
Note that $\mathcal{F}_{s} = O(1)$ and $e^{-i \varphi_{s}(0)} = O(1 / \sqrt{L \Gamma})$~\cite{Lee15_supp}. Therefore, Eq.~(\ref{Bscaling}) and Eq.~(\ref{S-scaling1}) yields
\begin{align}\label{S-scaling2}
\langle E_{i} | S_{\mathrm{imp}}^{-} | E_{j} \rangle = \begin{cases} O(1 / \sqrt{L \Gamma}) &\quad \text{for }\delta \Gamma \ll (1 / L) \ll \Gamma \\  O(1 / L \sqrt{\Gamma \delta \Gamma}) &\quad \text{for }(1/L) \ll \Gamma \: \& \: \delta \Gamma \end{cases}.
\end{align}
Now we replace $1/L$ by $E$ and identify $\Gamma$ with the Kondo temperature $T_{\mathrm{K}}$.
Since the crossover temperature is $T^{*} \propto \nu^{2} |\delta \lambda_{\perp}|^{2} T_{\mathrm{K}}$
and $\nu \sim 1 / a$,
we obtain $\delta \Gamma \sim T^{*}/(\nu T_\text{K})$.
By using these replacements, we find that
Eq.~(\ref{Szscaling2}) and Eq.~(\ref{S-scaling2}) are written as Eq.~(14) of the main text.


\begin{thebibliography}{99}
\bibitem{Singlet1} \HS{The representation $(|{\uparrow} \rangle \otimes |\phi_{i {\uparrow}} \rangle + |{\downarrow} \rangle \otimes |\phi_{i {\downarrow}} \rangle)/\sqrt{2}$ does not necessarily imply a spin-triplet state, as a sign factor $(-1)$ can be absorbed in the states $|\phi_{i {\uparrow}} \rangle$ or $ |\phi_{i {\downarrow}} \rangle$; \DHK{the representation can describe a Kondo singlet state}. $|\phi_{i \uparrow} \rangle$, $|\phi_{i \downarrow} \rangle$, and $|\delta E_{i} \rangle$ have spin-$z$ quantum numbers $S_{z}^{(i)} - 1/2$, $S_{z}^{(i)} + 1/2$, and $S_{z}^{(i)}$, respectively, where $S_{z}^{(i)}$ is the spin-$z$ quantum number of $|E_{i} \rangle$.}
\end{thebibliography}
\end{document}